\documentstyle[psfig,12pt]{article}
\def\lsim{\stackrel{<}{{}_\sim}}
\def\gsim{\stackrel{>}{{}_\sim}}
\def\be{\begin{equation}}       
\def\ee{\end{equation}}
\def\bear{\be\begin{array}}      
\def\eear{\end{array}\ee}
\def\bea{\begin{eqnarray}}
\def\eea{\end{eqnarray}}

\def\ie{{\it i.e.}}
\def\etal{{\it et al.}}
\def\half{{\textstyle{1 \over 2}}}

\def\quarter{{\textstyle{1 \over 4}}}

\def\bold#1{\setbox0=\hbox{$#1$}%
     \kern-.025em\copy0\kern-\wd0
     \kern.05em\copy0\kern-\wd0
     \kern-.025em\raise.0433em\box0 }

\catcode`\@=11  

\@addtoreset{equation}{section}

\def\theequation{\ksection.\arabic{equation}}

\begin{document}

\baselineskip 18pt

\newcommand{\sheptitle}
{\Large Radiative Corrections to Chargino Production in 
Electron-Positron Collisions}

\newcommand{\shepauthor}
{Marco A. D\'\i az${}^1$, Steve F. King${}^{2*}$, and Douglas A. Ross${}^3$}

\newcommand{\shepaddress}
{${}^1$ Departamento de F\'\i sica Te\'orica, IFIC--CSIC, Universidad de 
Valencia\\ 
Burjassot, Valencia 46100, Spain \\
${}^2$ CERN, Theory Division, CH--1211 Geneva, Switzerland\\
${}^3$ Department of Physics and Astronomy, University of Southampton\\
Southampton, SO17 1BJ, U.K.}

\newcommand{\shepabstract}
{
We discuss the one-loop radiative corrections to the reaction
$\sigma (e^+e^-\rightarrow \tilde{\chi}^+_a \tilde{\chi}^-_b)$, 
for $a,b=1,2$ where $\tilde{\chi}^{\pm}_{1,2}$ are the charginos 
of the minimal supersymmetric standard model. We calculate the leading 
one loop radiative corrections involving loops of top, stop, bottom and 
sbottom quarks, working in the $\overline{MS}$ scheme. At LEP2 we find
positive radiative corrections typically of $10\%$ to $15\%$ and with
a maximum value of approximately $30\%$ if the squark mass parameters
are of the order of 1 TeV. If $\sqrt{s}=500$ GeV we find smaller
corrections but they can be also negative, with extreme values of
$13\%$ and $-4\%$. For a center of mass given by $\sqrt{s}=2$ TeV
we find larger corrections, with typical values between $20\%$ and 
$-20\%$.
}

\begin{titlepage}
\begin{flushright}
CERN-TH/97-313\\
SHEP--97/23\\
FTUV/97--47\\
IFIC/97--78\\
hep-ph/9711307\\
\end{flushright}
\vspace{.1in}
\begin{center}
{\large{\bf \sheptitle}}
\bigskip \\ \bigskip\shepauthor\bigskip \\ 
\mbox{} \\ {\it \shepaddress} \\ \vspace{.5in}
{\bf Abstract} \bigskip \end{center} \setcounter{page}{0}
\shepabstract\\
\begin{flushleft}
CERN-TH/97-313\\
\today
\end{flushleft}

\noindent${}^*$ On leave of absence from $^3$.
\end{titlepage}

\section{Introduction}

The $e^+e^-$ colliders such as LEP
provide a clean environment for searching for the charginos and
neutralinos predicted by the
minimal supersymmetric standard model (MSSM) \cite{MSSMrep}.
Several authors have considered the production of neutralinos
and charginos at $e^+e^-$ colliders at the $Z$ pole \cite{prodLEP}
and beyond \cite{prodLEPII}, as well as its
decay modes \cite{decays}. From an accurate measurement of the
chargino production cross-section, much information could be obtained 
about the MSSM \cite{Baer, us}.

Experimental searches for charginos at LEP2 have been negative so far, 
and lower bounds on the lightest chargino mass have been set. The bound 
depends mainly on the sneutrino mass and the mass difference between the 
chargino and the LSP $\Delta m=m_{\chi_1}-m_{\chi^0_1}$.
ALEPH has found that $m_{\chi_1}>85$ GeV for $m_{\tilde\nu_e}>200$ GeV 
\cite{ALEPH}. DELPHI's bound corresponds
to $m_{\chi_1}>84.3$ GeV for $m_{\tilde\nu_e}>300$ GeV and 
$\Delta m>10$ GeV \cite{DELPHI}. A lower bound of $m_{\chi_1}>85.5$ GeV
was found by L3 for $m_{\tilde\nu_e}>300$ GeV \cite{L3}.
Finally, OPAL has found that if $\Delta m>10$ GeV then 
$m_{\chi_1}>84.5$ GeV if $m_0>1$ TeV and $m_{\chi_1}>65.7$ GeV for 
the smallest $m_0$ compatible with current limits on sneutrino
and slepton masses \cite{OPAL}.

Given the importance of an accurate measurement of the chargino production
cross-section in $e^+e^-$ experiments, it is clearly important to
be able to calculate this cross-section as accurately as possible.
Although this cross-section does not contain any coloured particles,
and so is immune to QCD corrections, there are other radiative corrections
which, as we shall see, may give large corrections to the cross-section
of order 10\%. Although electroweak corrections may be expected
to give corrections of order 1\%, there are additional radiative corrections
coming from loops of top and bottom quarks and squarks which are important
due to the large Yukawa couplings of these heavy quarks and squarks,
and it is these corrections which form the subject of the present paper.
Although radiative corrections to chargino masses 
have been considered
\cite{pierce}, the radiative corrections to chargino production
in $e^+e^-$ experiments has not so been considered in the literature.

The layout of this paper is as follows.
In Section 2 we discuss the tree level amplitude and outline the approach to
radiative corrections which we follow. In Section 3 we introduce 
convenient form factors which enable the calculation to be organized
in terms of its possible Lorentz invariant structures, and express the
square of the amplitude in terms of the form factors.
In section 4 we describe how the different diagrams contribute to the
form factors, relegating many of the details to a series of Appendices
where the Feynman rules and Passarino--Veltman (PV) functions 
are also summarised. Section 5 contains the numerical results,
and in Section 6 we give our conclusions.

\section{Tree Level Amplitude and Its One--Loop Renormalization}

We consider the pair production of charginos with momenta $k_1$ and $k_2$
in electron-positron scattering with incoming momenta $p_1$ and $p_2$:
\begin{equation}
e^+(p_2)+e^-(p_1)\rightarrow \tilde{\chi}^+_b(k_2) +\tilde{\chi}^-_a(k_1)
\end{equation}
In the MSSM charginos can be produced in the s--channel with intermediate
$Z$--bosons and photons, and in the t--channel with an intermediate 
electron--type sneutrino. We denote these amplitudes $M_Z^0$, 
$M_{\gamma}^0$, and $M_{\tilde\nu}^0$, where the superscript 0 indicates 
that the amplitude is at tree level. These amplitudes correspond to the 
diagrams in Fig.~\ref{ZGSneu1lAmplitudes} with the shaded blobs replaced 
by the lowest order tree-level vertices.

The tree level $Z$ amplitude can be written as
\begin{equation}
M_Z^0=\Big[\overline{v}(p_2)i{\cal G}_{Zee}^{0,\mu}u(p_1)\Big]
P_Z^{\mu\nu}(p^2)
\Big[\overline{u}(k_2)i{\cal G}_{Z\chi\chi}^{0,ab\nu}v(k_1)\Big]
\label{eq:MZtree}
\end{equation}
where
\begin{equation}
P_Z^{\mu\nu}(p^2)={{-ig^{\mu\nu}}\over{p^2-m_Z^2-im_Z\Gamma_Z}}
\label{eq:Zprop}
\end{equation}
is the $Z$--boson propagator in the Feynman gauge, with $m_Z$ its mass,
$\Gamma_Z$ its total width, and 
$p^2=(p_1+p_2)^2=(k_1+k_2)^2=s$
is the square of the center of mass energy.
The tree level $Ze^+e^-$ vertex function is
\begin{equation}
{\cal G}_{Zee}^{0,\mu}=-{g\over{2c_W}}\gamma^{\mu}(g_V^e-g_A^e\gamma_5)
\label{eq:ZeeTreeVertex}
\end{equation}
where $g$ is the $SU(2)$ gauge coupling, $s_W=\sin \theta_W$,
$c_W=\cos \theta_W$ where $\theta_W$ is
the weak angle, and $g_V^e=-1/2+2s_W^2$ and $g_A^e=-1/2$.

Similarly, the $Z\tilde\chi^+_b\tilde\chi^-_a$ vertices are given by
\begin{equation}
{\cal G}_{Z\chi\chi}^{0,ab\nu}={g\over{2c_W}}\gamma^{\nu}\Big[
O'^L_{ab}(1-\gamma_5)+O'^R_{ab}(1+\gamma_5)\Big]
\label{eq:ZchichiTreeVertex}
\end{equation}
where the couplings $O'^L_{ab}$ and $O'^R_{ab}$ are related to the 
matrices which diagonalize the chargino mass matrix, and are defined 
in Appendix A.

The tree level photon amplitude can be written as
\begin{equation}
M_{\gamma}^0=\Big[\overline{v}(p_2)i{\cal G}_{\gamma ee}^{0,\mu}u(p_1)\Big]
P_{\gamma}^{\mu\nu}(p^2)
\Big[\overline{u}(k_2)i{\cal G}_{\gamma\chi\chi}^{0,ab\nu}v(k_1)\Big]
\label{eq:MGtree}
\end{equation}
where the photon propagator in the Feynman gauge is
\begin{equation}
P_{\gamma}^{\mu\nu}(p^2)={{-ig^{\mu\nu}}\over{p^2}}
\label{eq:Gprop}
\end{equation}
The photon tree level vertices are very simple:
\begin{equation}
{\cal G}_{\gamma ee}^{0,\mu}=e\gamma^{\mu}\,,\qquad
{\cal G}_{\gamma\chi\chi}^{0,ab\nu}=-e\gamma^{\nu}\delta_{ab}
\label{eq:GeeGchichiVertex}
\end{equation}
where $e=|e|$ and the factor of $q_e=-1$ for the electron charge has 
been used.

Finally, the tree level sneutrino amplitude is
\begin{equation}
M_{\tilde\nu}^0=
\Big[\overline{v}(p_2)i{\cal G}_{\tilde\nu_ee\chi}^{0,+b}
\overline{u}^T(k_2)\Big]
P_{\tilde\nu_e}(q^2)
\Big[v^T(k_1)i{\cal G}_{\tilde\nu_ee\chi}^{0,-a}u(p_1)\Big]
\label{eq:MSneTree}
\end{equation}
where
\begin{equation}
P_{\tilde\nu_e}(q^2)={i\over{q^2-m_{\tilde\nu_e}^2}}
\label{eq:SnuProp}
\end{equation}
is the sneutrino propagator, $m_{\tilde\nu_e}$ is its mass and 
$q^2=(k_2-p_2)^2=(p_1-k_1)^2=t$ is the squared of the t--channel momentum 
transferred. The sneutrino vertices are at tree level given by
\begin{equation}
{\cal G}_{\tilde\nu_ee\chi}^{0,+b}=-{g\over 2}V_{b1}(1+\gamma_5)C
\,,\qquad
{\cal G}_{\tilde\nu_ee\chi}^{0,-a}={g\over 2}C^{-1}V_{a1}^*(1-\gamma_5)
\label{eq:SneTreeVertex}
\end{equation}
Here, $C$ is the charge conjugation matrix and $V$ is one of the
diagonalization matrices of the chargino mass matrix (see Appendix A).

Divergent diagrams are regularized using dimensional regularization. 
Therefore, the divergences are contained in the parameter
\begin{equation}
\Delta={2\over{4-n}}+\ln 4\pi-\gamma_E,
\label{eq:Delta}
\end{equation}
where $n$ is the number of space--time dimensions, and $\gamma_E$ is the 
Euler's constant. In every divergent graph, the term $\Delta$ is always
accompanied by the term $\ln Q^2$, where $Q$ is an arbitrary mass scale
introduced by dimensional regularization. The renormalization scheme we 
use here is the $\overline{MS}$. In this scheme, the counterterm is
fixed in such a way that cancels only the
terms proportional to $\Delta$. As a consequence, one--loop
corrections (diagrams plus counterterms) to 1PI Green's functions
become finite but remain 
scale dependent. In order to get a physical scattering amplitude, \ie,
independent of the scale $Q$, the tree level parameters are promoted to 
running parameters. This implicit scale dependence of the tree level
parameters cancels the explicit scale dependence of the one--loop 
corrections to the scattering amplitude.

We work in the approximation where only top and bottom quarks and squarks
are considered in the loops. These corrections are in general enhanced by 
logarithms of large mass ratios and by Yukawa couplings, whereas other 
corrections are genuinely of order $\alpha_W$ and therefore negligible.
This implies, for example, that the electron--positron vertices, 
${\cal G}_{Zee}^{\mu}$ and ${\cal G}_{\gamma ee}^{\mu}$, do not receive 
triangular corrections, and the tree level vertex can be identify with 
the one--loop renormalized vertex. In the following section we detail all 
these one--loop corrections to the chargino pair production.

\section{Squared Amplitudes in Terms of Form Factors}

\begin{figure}
\centerline{\protect\hbox{\psfig{file=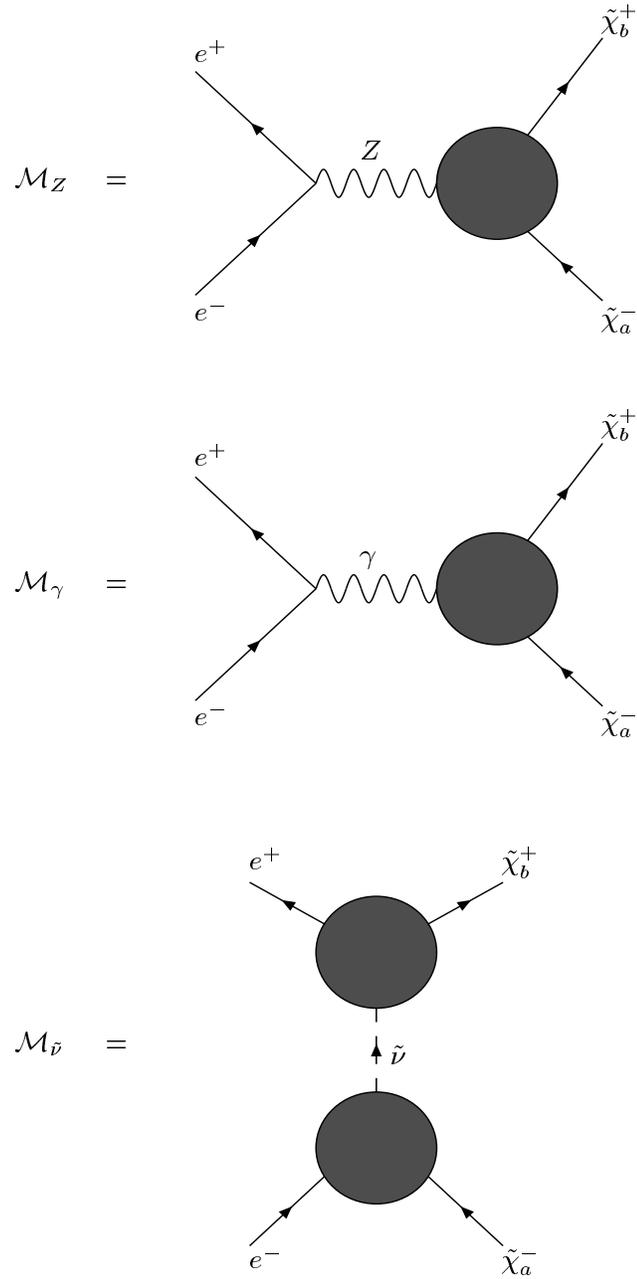,height=17cm,width=0.55\textwidth}}}
\caption{One--loop renormalized $M_Z$, $M_{\gamma}$ and 
$M_{\tilde\nu_e}$ amplitudes in the approximation where only top and
bottom quarks and squarks are considered inside the loops.} 
\label{ZGSneu1lAmplitudes} 
\end{figure} 
In the presence of radiative corrections,
the amplitude for $e^+e^-\rightarrow \tilde{\chi}^+_b\tilde{\chi}^-_a$
may be expressed as the sum of three amplitudes $M_Z$, $M_{\gamma}$,  
$M_{\tilde{\nu}}$ as shown in Fig.~\ref{ZGSneu1lAmplitudes}. The shaded bubbles
in that figure are one--loop renormalized total vertex functions defined as
$i{\cal G}_{Z\chi\chi}^{ab}$, $i{\cal G}_{\gamma\chi\chi}^{ab}$, 
$i{\cal G}_{\tilde{\nu_e}e\chi}^{+b}$, and
$i{\cal G}_{\tilde{\nu_e}e\chi}^{-a}$.
In the total vertex functions we include the tree level vertex, the
one--particle irreducible vertex diagrams plus the vertex counterterm, and
the one--particle reducible vertex diagrams plus their counterterms.
Although the detailed expressions for the
total vertex functions is quite complicated, by exploiting the possible
Lorentz structures of the diagrams it is possible to
express them in terms of just a few form factors.   
 
We define the Z form factors as follows:
\begin{equation}
{\cal G}_{Z\chi\chi}^{ab}
\equiv (1+\gamma_5)\left(F_{Z0}^+\gamma^{\mu}+F_{Z1}^+
k_1^{\mu}+F_{Z2}^+k_2^{\mu}\right)+(1-\gamma_5)\left(F_{Z0}^-
\gamma^{\mu}+F_{Z1}^-k_1^{\mu}+F_{Z2}^-k_2^{\mu}\right)
\label{eq:FormZXX}
\end{equation}
Similarly we define the photon form factors as follows:
\begin{equation}
{\cal G}_{\gamma\chi\chi}^{ab}
\equiv (1+\gamma_5)\left(F_{\gamma0}^+\gamma^{\mu}+
F_{\gamma1}^+k_1^{\mu}+F_{\gamma2}^+k_2^{\mu}\right)+(1-\gamma_5)
\left(F_{\gamma0}^-\gamma^{\mu}+F_{\gamma1}^-k_1^{\mu}+F_{\gamma2}^-
k_2^{\mu}\right)
\label{eq:FormPhXX}
\end{equation}
Sneutrino total vertex functions are simpler. 
There are two total sneutrino vertex functions denoted by 
${\cal G}_{\tilde{\nu_e}e\chi}^{\pm a}$, and each of them can be express 
in terms of a single form factor. For the upper vertex we have:
\begin{equation}
{\cal G}_{\tilde{\nu_e}e\chi}^{+b}
\equiv 
(1+\gamma_5)CF_{\tilde{\nu_e}}^{+}
\end{equation}
For the lower vertex we have:
\begin{equation}
{\cal G}_{\tilde{\nu_e}e\chi}^{-a}
\equiv 
C^{-1}(1-\gamma_5)F_{\tilde{\nu_e}}^{-}
\end{equation}

Having expressed the vertices $Z\chi_b^+\chi_a^-$,
$\gamma\chi_b^+\chi_a^-$, $\tilde{\nu}e^+\chi_b^+$,
and $\tilde{\nu}e^-\chi_a^-$ in terms of form factors,
it is a relatively straightforward task to write down the one-loop
amplitudes in terms of the form factors.

The $Z$ amplitude squared is
\begin{eqnarray}
\langle|M_Z|^2\rangle&=&{{g^2}\over{c_W^2}}{{s^2}\over{(s-m_Z^2)^2+
m_Z^2\Gamma_Z^2}}\Bigg\{
\nonumber\\
&&2g_V^eg_A^e\left(|F_{Z0}^-|^2-|F_{Z0}^+|^2\right)R+\left(g_V^{e2}
+g_A^{e2}\right)\bigg\{4m_{\chi_a}m_{\chi_b}{\mathrm Re}(F_{Z0}^+
F_{Z0}^{-*})/s
\nonumber\\
&&+{\textstyle{1\over 2}}\left(|F_{Z0}^+|^2+|F_{Z0}^-|^2\right)\left[
1+R^2-(m_{\chi_a}^2-m_{\chi_b}^2)^2/s^2\right]
\nonumber\\
&&+{\textstyle{1\over 2}}\bigg[{\mathrm Re}\left[(m_{\chi_a}F_{Z0}^-+
m_{\chi_b}F_{Z0}^+)(F_{Z1}^--F_{Z2}^-)^*\right]
\nonumber\\
&&\qquad+{\mathrm Re}\left[(m_{\chi_a}F_{Z0}^++m_{\chi_b}F_{Z0}^-)
(F_{Z1}^+-F_{Z2}^+)^*\right]
\nonumber\\
&&\qquad+{\textstyle{1\over4}}(s-m_{\chi_a}^2-m_{\chi_b}^2)\left(
|F_{Z1}^--F_{Z2}^-|^2+|F_{Z1}^+-F_{Z2}^+|^2\right)
\nonumber\\
&&\qquad-m_{\chi_a}m_{\chi_b}{\mathrm Re}\left[(F_{Z1}^--F_{Z2}^-)
(F_{Z1}^+-F_{Z2}^+)^*\right]
\nonumber\\
&&\!\!\qquad\bigg]\left[1-R^2-2(m_{\chi_a}^2+m_{\chi_b}^2)/s+
(m_{\chi_a}^2-m_{\chi_b}^2)^2/s^2\right]\bigg\}\Bigg\}
\label{eq:MZSquare}
\end{eqnarray}
where 
\begin{equation}
R=\lambda^{1/2}(1,m_{\chi_a}^2/s,m_{\chi_b}^2/s)\cos\theta\,,
\label{eq:Rdifmass}
\end{equation}
$\lambda(a,b,c)=(a+b-c)^2-4ab$, and $\theta$ is the angle between the 
electron and chargino lines.
Note that we have already sum over final spin configurations and taken
the average of initial spin configurations. The photon amplitude squared
is given by
\begin{eqnarray}
\langle|M_{\gamma}|^2\rangle&=&2e^2\Bigg\{8m_{\chi_a}m_{\chi_b}
{\mathrm Re}(F_{\gamma0}^+F_{\gamma0}^{-*})/s
\nonumber\\
&&+\left(|F_{\gamma0}^+|^2+|F_{\gamma0}^-|^2\right)\left[
1+R^2-(m_{\chi_a}^2-m_{\chi_b}^2)^2/s^2\right]
\nonumber\\
&&+\bigg[{\mathrm Re}\left[(m_{\chi_a}F_{\gamma0}^-+m_{\chi_b}
F_{\gamma0}^+)(F_{\gamma1}^--F_{\gamma2}^-)^*\right]
\nonumber\\
&&\qquad+{\mathrm Re}\left[(m_{\chi_a}F_{\gamma0}^++m_{\chi_b}
F_{\gamma0}^-)(F_{\gamma1}^+-F_{\gamma2}^+)^*\right]
\nonumber\\
&&\qquad+{\textstyle{1\over4}}(s-m_{\chi_a}^2-m_{\chi_b}^2)\left(
|F_{\gamma1}^--F_{\gamma2}^-|^2+|F_{\gamma1}^+-F_{\gamma2}^+|^2\right)
\nonumber\\
&&\qquad-m_{\chi_a}m_{\chi_b}{\mathrm Re}\left[(F_{\gamma1}^--F_{\gamma2}^-)
(F_{\gamma1}^+-F_{\gamma2}^+)^*\right]
\nonumber\\
&&\quad\bigg]\left[1-R^2-2(m_{\chi_a}^2+m_{\chi_b}^2)/s+
(m_{\chi_a}^2-m_{\chi_b}^2)^2/s^2\right]\Bigg\}
\label{eq:MPhotSquare}
\end{eqnarray}
and the sneutrino amplitude squared is
\begin{equation}
\langle|M_{\tilde\nu_e}|^2\rangle=
|F_{\tilde\nu_e}^+|^2|F_{\tilde\nu_e}^-|^2
{{s^2}\over{(t-m_{\tilde\nu_e}^2)^2}}\left[(1-R)^2-(m_{\chi_a}^2
-m_{\chi_b}^2)^2/s^2\right]
\label{eq:MSneuSquare}
\end{equation}
The total amplitude squared has three interferences. We start with
the $Z$--photon interference
\begin{eqnarray}
&&2{\mathrm Re}\langle M_ZM_{\gamma}^*\rangle\;\>=\;\>{{eg}\over{c_W}}
{\mathrm Re}{s\over{s-m_Z^2-im_Z\Gamma_Z}}\Bigg\{
4g_A^e(F_{Z0}^-F_{\gamma0}^{-*}-F_{Z0}^+F_{\gamma0}^{+*})+g_V^e\bigg\{
\nonumber\\
&&\qquad\;8m_{\chi_a}m_{\chi_b}(F_{Z0}^+
F_{\gamma0}^{-*}+F_{Z0}^-F_{\gamma0}^{+*})/s
\nonumber\\
&&\qquad+2(F_{Z0}^+F_{\gamma0}^{+*}+F_{Z0}^-F_{\gamma0}^{-*})
\left[1+R^2-(m_{\chi_a}^2-m_{\chi_b}^2)^2/s^2\right]
\nonumber\\
&&\qquad+\bigg[(m_{\chi_a}F_{Z0}^-+m_{\chi_b}F_{Z0}^-)(
F_{\gamma1}^--F_{\gamma2}^-)^*+(m_{\chi_a}F_{Z0}^++m_{\chi_b}F_{Z0}^-)
(F_{\gamma1}^+-F_{\gamma2}^+)^*
\nonumber\\
&&\qquad\quad+(F_{Z1}^--F_{Z2}^-)(m_{\chi_a}F_{\gamma0}^{-*}+m_{\chi_b}
F_{\gamma0}^{+*})+(F_{Z1}^+-F_{Z2}^+)(m_{\chi_a}F_{\gamma0}^{+*}+
m_{\chi_b}F_{\gamma0}^{-*})
\nonumber\\
&&\qquad\quad+{\textstyle{1\over2}}(s-m_{\chi_a}^2-m_{\chi_b}^2)\left[
(F_{Z1}^--F_{Z2}^-)(F_{\gamma1}^--F_{\gamma2}^-)^*+(F_{Z1}^+-
F_{Z2}^+)(F_{\gamma1}^+-F_{\gamma2}^+)^*\right]
\nonumber\\
&&\qquad\quad-m_{\chi_a}m_{\chi_b}\left[(F_{Z1}^--F_{Z2}^-)
(F_{\gamma1}^+-F_{\gamma2}^+)^*+(F_{Z1}^+-F_{Z2}^+)(F_{\gamma1}^--
F_{\gamma2}^-)^*\right]
\nonumber\\
&&\qquad\quad\bigg]\left[1-R^2-2(m_{\chi_a}^2+m_{\chi_b}^2)/s+
(m_{\chi_a}^2-m_{\chi_b}^2)^2/s^2\right]\bigg\}\Bigg\}
\label{eq:ZPhotInt}
\end{eqnarray}
The $Z$-sneutrino interference is
\begin{eqnarray}
2{\mathrm Re}\langle M_ZM_{\tilde\nu_e}^*\rangle&=&{{g(g_V^e+g_A^e)
}\over{c_W}}{\mathrm Re}{{s^2}\over{(s-m_Z^2-im_Z\Gamma_Z)
(t-m_{\tilde\nu_e}^2)}}
F_{\tilde{\nu_e}}^{+}F_{\tilde{\nu_e}}^{-}\bigg\{
\nonumber\\
&&4F_{Z0}^-m_{\chi_a}m_{\chi_b}/s+F_{Z0}^+\left[(1-R)^2-(m_{\chi_a}^2
-m_{\chi_b}^2)^2/s^2\right]
\nonumber\\
&&+{\textstyle{1\over2}}\Big[m_{\chi_a}(F_{Z1}^+-F_{Z2}^+)+m_{\chi_b}
(F_{Z1}^--F_{Z2}^-)
\nonumber\\
&&\quad\;\,\Big]\left[1-R^2-2(m_{\chi_a}^2+m_{\chi_b}^2)/s+
(m_{\chi_a}^2-m_{\chi_b}^2)^2/s^2\right]\bigg\}
\label{eq:ZSneuInt}
\end{eqnarray}
and the photon--sneutrino interference is given by
\begin{eqnarray}
2{\mathrm Re}\langle M_{\gamma}M_{\tilde\nu_e}^*\rangle&=&
2e{s\over{(t-m_{\tilde\nu_e}^2)}}{\mathrm Re}\,
F_{\tilde{\nu_e}}^{+}F_{\tilde{\nu_e}}^{-}\bigg\{
\nonumber\\
&&4F_{\gamma0}^-m_{\chi_a}m_{\chi_b}/s+F_{\gamma0}^+\left[(1-R)^2-
(m_{\chi_a}^2-m_{\chi_b}^2)^2/s^2\right]
\nonumber\\
&&+{\textstyle{1\over2}}\Big[m_{\chi_a}(F_{\gamma1}^+-F_{\gamma2}^+)+
m_{\chi_b}(F_{\gamma1}^--F_{\gamma2}^-)
\nonumber\\
&&\quad\;\,\Big]\left[1-R^2-2(m_{\chi_a}^2+m_{\chi_b}^2)/s+
(m_{\chi_a}^2-m_{\chi_b}^2)^2/s^2\right]\bigg\}
\label{eq:PhotSneuInt}
\end{eqnarray}
Now we just have to calculate the one--loop contributions to each form
factor, and this is done in the next section.

\section{Loop Contributions to Form Factors}

In the previous section we defined the total vertex functions 
$i{\cal G}_{Z\chi\chi}^{ab}$, $i{\cal G}_{\gamma\chi\chi}^{ab}$,
and ${\cal G}_{\tilde{\nu_e}e\chi}^{\pm a}$ expressed in terms of form 
factors $F_{Zi}^{\pm}$, $F_{\gamma i}^{\pm}$, with $i=0,1,2$,
and $F_{\tilde\nu_e}^{\pm}$.
In this section we give details about the contributions from the different 
diagrams to the form factors defined in the previous section. We divide the
diagrams into one--particle irreducible triangular diagrams and 
one--particle reducible diagrams composed by gauge
boson two--point functions, chargino mixing, and chargino wavefunction
renormalization (see Appendix C).

\begin{figure}
\centerline{\protect\hbox{\psfig{file=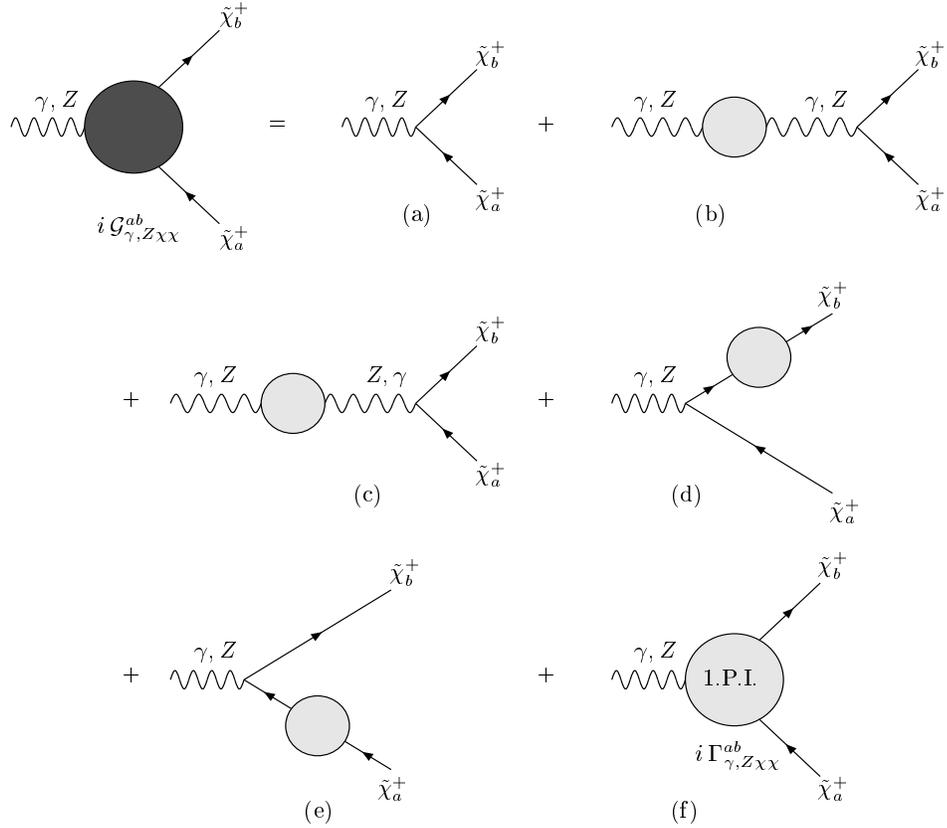,height=11cm,width=0.8\textwidth}}}
\caption{One--loop renormalized $\gamma\chi^+_b\chi^-_a$ and 
$Z\chi^+_b\chi^-_a$ vertex functions.} 
\label{GZchacha1lvertex} 
\end{figure}

The sum of the one-particle-irreducible vertex diagrams 
$i{\Gamma}_{Z\chi\chi}^{ab}$ includes one-loop triangle diagrams with 
internal top, bottom, stop, and sbottom quarks. Each of these 
graphs in terms of Passarino--Veltman (PV) functions \cite{VP} are given 
in Appendix C. 
{}From there one can read off the contribution of each graph to the form
factors $F_{Zi}^{\pm}$, $i=0,1,2$, defined in eq.~(\ref{eq:FormZXX}). 
This is particularly simple because the one-particle-irreducible vertex 
function $i{\Gamma}_{Z\chi\chi}^{ab}$ has the same Lorentz structure 
as the total vertex function $i{\cal G}_{Z\chi\chi}^{ab}$ in 
eq.~(\ref{eq:FormZXX}). These contributions are graphically represented 
by the diagram in Fig.~\ref{GZchacha1lvertex}f, with an external 
$Z$-boson. 

The sum of all one-particle-irreducible vertex diagrams 
$i{\Gamma}_{\gamma\chi\chi}^{ab}$ are treated in a completely analogous 
way. These graphs are also given in Appendix C, and from those 
expressions one can read off the corresponding contributions to the form factors 
$F_{\gamma i}^{\pm}$ in the total vertex function 
$i{\cal G}_{\gamma\chi\chi}^{ab}$. These contributions are also 
represented by the diagram in Fig.~\ref{GZchacha1lvertex}f, but with an 
external photon attached to it. 

There are no contributions from triangular graphs to the form factors 
$F_{\tilde\nu_e}^{\pm}$ in the total vertex functions 
${\cal G}_{\tilde{\nu_e}e\chi}^{\pm a}$. In the following subsections 
we analyse the one-particle-reducible graphs.

\subsection{Gauge Bosons Two--Point Functions}

The contribution from the gauge boson two--point functions is
displayed in Fig.~\ref{GZchacha1lvertex}b and \ref{GZchacha1lvertex}c. 
The general structure of the gauge boson two--point functions is
\begin{equation}
\Sigma_{GG}^{\mu\nu}(p^2)=
i[A_{GG}(p^2)g^{\mu\nu}+B_{GG}(p^2)p^{\mu}p^{\nu}]
\label{eq:GauGau2Funct}
\end{equation}
where $p$ is the external momentum and 
$GG=ZZ$, $\gamma\gamma$, or $Z\gamma$. The functions $A_{GG}$ and $B_{GG}$ 
depend on the external momentum squared 
$p^2$ and represent the one--loop contributions to the gauge
bosons self energies and mixing. For our purposes only $A_{GG}$ is 
relevant and these are displayed in Appendix C2. 
The gauge boson two--point functions attached to the external
chargino line give contributions to the forms factors $F_{Z0}^{\pm}$
and $F_{\gamma 0}^{\pm}$ only.

The $Z$--boson self energy contributes to $F_{Z0}^+$ and 
$F_{Z0}^-$ form factors in the following way
\begin{equation}
F_{Z0}^+={g\over{2c_W}}O'^L_{ab}{{\widetilde A_{ZZ}(p^2)}\over{p^2-m_Z^2}}
\,,\qquad
F_{Z0}^-={g\over{2c_W}}O'^R_{ab}{{\widetilde A_{ZZ}(p^2)}\over{p^2-m_Z^2}}
\,,\label{eq:ZZtoform}
\end{equation}
where $p^2=s$, the center--of--mass energy, and
the tilde in $\widetilde A$ represent the self energy plus its 
counterterm, \ie, $\widetilde A$ is the finite two--point function.

Another contribution to the same form factors come from the $Z-\gamma$
mixing:
\begin{equation}
F_{Z0}^+=-{e\over 2}\delta_{ab}{{\widetilde A_{Z\gamma}(p^2)}\over{p^2}}
\,,\qquad
F_{Z0}^-=-{e\over 2}\delta_{ab}{{\widetilde A_{Z\gamma}(p^2)}\over{p^2}}
\,,\label{eq:ZGammatoform}
\end{equation}
where we take $e$ to be positive.

In a similar way, the $F_{\gamma 0}^+$ and $F_{\gamma 0}^-$ form factors
receive the following contributions from the photon self energy
\begin{equation}
F_{\gamma 0}^+=-{e\over 2}\delta_{ab}{{\widetilde A_{\gamma\gamma}(p^2)}
\over{p^2}}\,,\qquad
F_{\gamma 0}^-=-{e\over 2}\delta_{ab}{{\widetilde A_{\gamma\gamma}(p^2)}
\over{p^2}}\,,
\label{eq:GammaGammatoform}
\end{equation}
and from the $\gamma-Z$ mixing
\begin{equation}
F_{\gamma 0}^+={g\over{2c_W}}O'^L_{ab}{{\widetilde A_{\gamma Z}(p^2)}
\over{p^2-m_Z^2}}\,,\qquad
F_{\gamma 0}^-={g\over{2c_W}}O'^R_{ab}{{\widetilde A_{\gamma Z}(p^2)}
\over{p^2-m_Z^2}}\,.
\label{eq:GammaZtoform}
\end{equation}
There is no contribution from gauge two--point functions to the 
${\cal G}_{\tilde{\nu_e}e\chi}^{\pm a}$ total vertex functions.

\subsection{Chargino Mixing}

Chargino mixing graphs contribute to the total vertices 
${\cal G}_{\gamma\chi\chi}^{ab}$ and ${\cal G}_{Z\chi\chi}^{ab}$ through
diagrams represented in Fig.~\ref{GZchacha1lvertex}d and 
\ref{GZchacha1lvertex}e.

The most general expression for the chargino two--point functions
at one--loop is
\begin{equation}
i\Sigma_{\chi\chi}^{ij}(p)=i\bigg\{
\Big[A_{ij}^+(p^2)(1+\gamma_5)+A_{ij}^-(p^2)(1-\gamma_5)\Big]+
\Big[B_{ij}^+(p^2)(1+\gamma_5)+B_{ij}^-(p^2)(1-\gamma_5)\Big]
p_{\mu}\gamma^{\mu}\bigg\}
\label{eq:ChaChaGeneral}
\end{equation}
and the contribution to the functions $A_{ij}^{\pm}$ and $B_{ij}^{\pm}$
from the different loops are given in Appendix C. Chargino mixings 
contribute to the form factors when the loop is attached to an 
external chargino. We start with the $Z\chi^+_b\chi^-_a$ vertex. If the
loop is attached to the external $\tilde\chi^+_b$ chargino, we find
the following contributions
\begin{eqnarray}
F_{Z0}^+&=&-{g\over{c_W}}O'^L_{ai}{1\over{m_{\chi_b}^2-m_{\chi_i}^2}}
\Big[m_{\chi_b}\widetilde A_{ib}^-+m_{\chi_i}\widetilde A_{ib}^++
m_{\chi_b}(m_{\chi_b}\widetilde B_{ib}^++m_{\chi_i}\widetilde B_{ib}^-)
\Big]
\nonumber\\
F_{Z0}^-&=&-{g\over{c_W}}O'^R_{ai}{1\over{m_{\chi_b}^2-m_{\chi_i}^2}}
\Big[m_{\chi_b}\widetilde A_{ib}^++m_{\chi_i}\widetilde A_{ib}^-+
m_{\chi_b}(m_{\chi_b}\widetilde B_{ib}^-+m_{\chi_i}\widetilde B_{ib}^+)
\Big]
\label{eq:bChaMixZForm}
\end{eqnarray}
where $i\ne b$. All the functions $\widetilde A^{\pm}(k_2^2)$ and
$\widetilde B^{\pm}(k_2^2)$ are evaluated at $k_2^2=m_{\chi_b}^2$, and
the tilde means that the corresponding function is finite.
If the one--loop graph is attached to the external $\tilde\chi^-_a$ 
chargino, then the contributions to the $F^{\pm}_{Z0}$ form factors are
\begin{eqnarray}
F_{Z0}^+&=&-{g\over{c_W}}O'^L_{bi}{1\over{m_{\chi_a}^2-m_{\chi_i}^2}}
\Big[m_{\chi_a}\widetilde A_{ai}^++m_{\chi_i}\widetilde A_{ai}^-+
m_{\chi_a}(m_{\chi_a}\widetilde B_{ai}^++m_{\chi_i}\widetilde B_{ai}^-)
\Big]
\nonumber\\
F_{Z0}^-&=&-{g\over{c_W}}O'^R_{bi}{1\over{m_{\chi_a}^2-m_{\chi_i}^2}}
\Big[m_{\chi_a}\widetilde A_{ai}^-+m_{\chi_i}\widetilde A_{ai}^++
m_{\chi_a}(m_{\chi_a}\widetilde B_{ai}^-+m_{\chi_i}\widetilde B_{ai}^+)
\Big]
\label{eq:aChaMixZForm}
\end{eqnarray}
where $i\ne a$ and all the functions $\widetilde A^{\pm}(k_1^2)$ and
$\widetilde B^{\pm}(k_1^2)$ are evaluated at $k_1^2=m_{\chi_a}^2$.

In a similar way, we find the contributions to the form factors associated
to the vertex $\gamma\chi^+_b\chi^-_a$. If the chargino mixing graph is 
attached to the external $\tilde\chi^+_b$ chargino, we find
\begin{eqnarray}
F_{\gamma 0}^+&=&{e\over{m_{\chi_b}^2-m_{\chi_a}^2}}
\Big[m_{\chi_b}\widetilde A_{ab}^-+m_{\chi_a}\widetilde A_{ab}^++
m_{\chi_b}(m_{\chi_b}\widetilde B_{ab}^++m_{\chi_a}\widetilde B_{ab}^-)
\Big]
\nonumber\\
F_{\gamma 0}^-&=&{e\over{m_{\chi_b}^2-m_{\chi_a}^2}}
\Big[m_{\chi_b}\widetilde A_{ab}^++m_{\chi_a}\widetilde A_{ab}^-+
m_{\chi_b}(m_{\chi_b}\widetilde B_{ab}^-+m_{\chi_a}\widetilde B_{ab}^+)
\Big]
\label{eq:bChaMixGForm}
\end{eqnarray}
where as before, all the functions $\widetilde A^{\pm}(k_2^2)$ and
$\widetilde B^{\pm}(k_2^2)$ are evaluated at $k_2^2=m_{\chi_b}^2$. 
The difference with respect to the previous case is that here we need 
$a\ne b$, otherwise this contribution is absent.

If the one--loop graph is attached to the external $\tilde\chi^-_a$ 
chargino, we get
\begin{eqnarray}
F_{\gamma 0}^+&=&{e\over{m_{\chi_a}^2-m_{\chi_b}^2}}
\Big[m_{\chi_a}\widetilde A_{ab}^++m_{\chi_b}\widetilde A_{ab}^-+
m_{\chi_a}(m_{\chi_a}\widetilde B_{ab}^++m_{\chi_b}\widetilde B_{ab}^-)
\Big]
\nonumber\\
F_{\gamma 0}^-&=&{e\over{m_{\chi_a}^2-m_{\chi_b}^2}}
\Big[m_{\chi_a}\widetilde A_{ab}^-+m_{\chi_b}\widetilde A_{ab}^++
m_{\chi_a}(m_{\chi_a}\widetilde B_{ab}^-+m_{\chi_b}\widetilde B_{ab}^+)
\Big]
\label{eq:aChaMixGForm}
\end{eqnarray}
where, again, we need $a\ne b$ for this contribution to be different 
from zero. As before, all the functions $\widetilde A^{\pm}(k_1^2)$ 
and $\widetilde B^{\pm}(k_1^2)$ are evaluated at $k_1^2=m_{\chi_a}^2$.

\begin{figure}
\centerline{\protect\hbox{\psfig{file=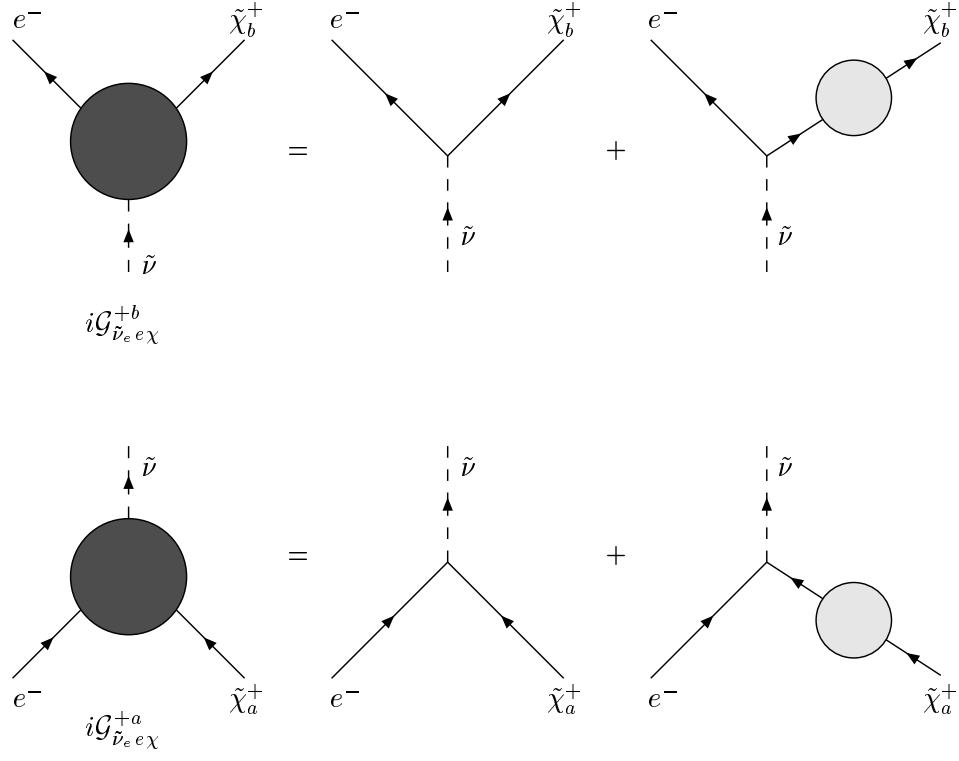,height=10cm,width=0.83\textwidth}}}
\caption{One--loop renormalized $e^-\tilde\nu_e\tilde\chi^+$ 
vertex functions.} 
\label{eSnueCha1lvertex} 
\end{figure} 
Finally, we consider the $e^+\tilde\nu_e\tilde\chi^+_b$ 
and $e^-\tilde\nu_e\tilde\chi^-_a$ vertices,
whose renormalization is represented in Fig.~\ref{eSnueCha1lvertex}.
If the one--loop graph is attached to the external $\tilde\chi^+_b$ 
chargino we find a contribution to the sneutrino form factor given by
\begin{equation}
F_{\tilde\nu_e}^+={{gV_{i1}}\over{m_{\chi_b}^2-m_{\chi_i}^2}}
\Big[m_{\chi_b}\widetilde A_{ib}^-+m_{\chi_i}\widetilde A_{ib}^++
m_{\chi_b}(m_{\chi_b}\widetilde B_{ib}^++m_{\chi_i}\widetilde B_{ib}^-)
\Big]
\label{eq:bChaMixSForm}
\end{equation}
with $i\ne b$ and the all the functions $\widetilde A^{\pm}(k_2^2)$ and
$\widetilde B^{\pm}(k_2^2)$ are evaluated at $k_2^2=m_{\chi_b}^2$.
If the one--loop graph is attached to the external $\tilde\chi^-_a$ 
chargino we get
\begin{equation}
F_{\tilde\nu_e}^-=-{{gV_{i1}^*}\over{m_{\chi_a}^2-m_{\chi_i}^2}}
\Big[m_{\chi_a}\widetilde A_{ai}^++m_{\chi_i}\widetilde A_{ai}^-+
m_{\chi_a}(m_{\chi_a}\widetilde B_{ai}^++m_{\chi_i}\widetilde B_{ai}^-)
\Big]
\label{eq:aChaMixSForm}
\end{equation}
with $i\ne a$ and the all the functions $\widetilde A^{\pm}(k_1^2)$ and
$\widetilde B^{\pm}(k_1^2)$ are evaluated at $k_1^2=m_{\chi_a}^2$.

\subsection{Chargino Masses and Wavefunction Renormalization}

The unrenormalized chargino two--point functions in 
eq.~(\ref{eq:ChaChaGeneral}) can be decomposed into two terms:
\begin{equation}
\Sigma_{\chi\chi}^{ij}(p)=\Sigma_{\chi\chi,1}^{ij}(p)
+\gamma_5\Sigma_{\chi\chi,5}^{ij}(p)
\label{eq:Sigma1y5}\,.
\end{equation}
The inverse propagator at one--loop is then obtained by adding this
self energy, previously renormalized according to the $\overline{MS}$
scheme, to the tree level propagator with the tree level mass promoted
to a running mass
\begin{equation}
\Gamma^{(2)}_{\chi\chi}(p)=p_{\mu}\gamma^{\mu}-m_{\chi_i}(Q)+
\widetilde\Sigma_{\chi\chi}^{ii}(p,Q)
\label{InvProp}
\end{equation}
The full propagator is essentially just the inverse of 
$\Gamma^{(2)}_{\chi\chi}$, so the physical pole mass is given by the zero 
of this function, in the limit where 
$p_{\mu}\gamma^{\mu}\longrightarrow m_{\chi_i}$, 
and may be found with the aid of the following equation:
\begin{equation}
\widetilde Z_{\chi_i}^{-1}\,
\overline{u}(p)\Big[p_{\mu}\gamma^{\mu}-m_{\chi_i}\Big]u(p)=
\overline{u}(p)\Big[p_{\mu}\gamma^{\mu}-m_{\chi_i}(Q)+
\widetilde\Sigma_{\chi\chi}^{ii}(p,Q)\Big]u(p)
\label{eq:MassRen}
\end{equation}
where $u$ and $\overline{u}$ are two (on-shell) spinors, $m_{\chi_i}$ 
is the pole mass of the chargino $\tilde\chi_i$, $m_{\chi_i}(Q)$ is its 
running mass, and $\widetilde\Sigma_{\chi\chi}^{ij}(p,Q)$ is the 
renormalized chargino two--point function in the $\overline{MS}$ scheme.
The quantity $\widetilde Z_{\chi_i}$ is the residual finite wavefunction
renormalization in the $\overline{MS}$ scheme, which account for the
fact that the residue of the $\overline{MS}$ propagator at the pole is 
not one, as we shall see later. $\widetilde Z_{\chi_i}$ corresponds to
the finite ratio of the infinite wavefunction 
renormalization constants in the $\overline{MS}$ scheme and the on-shell 
scheme. When renormalization is performed in the $\overline{MS}$ scheme, 
each external $\tilde\chi_i$ line has a factor of  
$(\widetilde Z_{\chi_i})^{-1/2}$ associated with it, according to 
the LSZ reduction formula. The renormalized
two--point function $\widetilde\Sigma_{\chi\chi}^{ij}(p,Q)$ is calculated 
simply by substracting the pole terms proportional to the regulator of 
dimensional regularization $\Delta$ defined in eq.~(\ref{eq:Delta}). Since 
$\overline{u}(p)\gamma_5u(p)=0$, only 
$\widetilde\Sigma_{\chi\chi,1}^{ij}(p)$ survives. From here
we deduce the relation between the pole and running chargino masses:
\begin{equation}    
\Delta m_{\chi_i} \equiv
m_{\chi_i}(Q)-m_{\chi_i}=\widetilde A_{ii}^+(m_{\chi_i}^2)+
\widetilde A_{ii}^-(m_{\chi_i}^2)+m_{\chi_i}\Big[
\widetilde B_{ii}^+(m_{\chi_i}^2)+\widetilde B_{ii}^-(m_{\chi_i}^2)\Big]
\label{eq:ChaPoleRunM}
\end{equation}
Note that since $Q$ is the subtraction point, the renormalized quantities
$\widetilde{A}^{\pm}$ and $\widetilde{B}^{\pm}$ are explicitly functions
of $Q$. We give all our results in terms of the physical mass, \ie, the pole
mass $m_{\chi_i}$.

In order to determine  $\widetilde Z_{\chi_i}$   we first
find  the one--particle reducible graph formed by
the sum of an infinite number of one--particle irreducible
two point functions $i\widetilde\Sigma_{\chi\chi}^{ii}(p,Q)$ 
connected in series:
\begin{eqnarray}
&&\overline{u}(p){i\over{p_{\mu}\gamma^{\mu}-m_{\chi_i}(Q)}}
\Bigg\{1+i\widetilde\Sigma_{\chi\chi}^{ii}(p,Q)
{i\over{p_{\mu}\gamma^{\mu}-m_{\chi_i}(Q)}}+...\Bigg\}u(p)=
\nonumber\\
&&\overline{u}(p){i\over{p_{\mu}\gamma^{\mu}-m_{\chi_i}(Q)+
\widetilde\Sigma_{\chi\chi}^{ii}(p,Q)}}u(p)=
\overline{u}(p){{i\widetilde Z_{\chi_i}}\over{p_{\mu}\gamma^{\mu}-
m_{\chi_i}}}u(p)
\label{eq:PropRen}
\end{eqnarray}
and the last equality in eq.~(\ref{eq:PropRen}) tell us that
after taking the limit $p^2\rightarrow m_{\chi_i}^2$ and 
$p_{\mu}\gamma^{\mu}u\rightarrow m_{\chi_i}u$, the effect
of the radiative corrections is to introduce a finite renormalization
factor given by
\begin{equation}
\widetilde Z_{\chi_i}=1-2m_{\chi_i}\bigg[\widetilde A'^+_{ii}+
\widetilde A'^-_{ii}+
m_{\chi_i}\Big(\widetilde B'^+_{ii}+\widetilde B'^-_{ii}\Big)\bigg]-
\widetilde B^+_{ii}-\widetilde B^-_{ii}
\label{eq:Zchi}
\end{equation}
where the prime indicate derivative with respect to the argument $p^2$,
and all the functions are evaluated at $p^2=m_{\chi_i}^2$.
Note that the terms proportional to $\gamma_5\Sigma_{\chi\chi,5}^{ij}(p)$
do not contribute either to the difference between the pole mass and the 
running mass, or to the wavefunction renormalisation factor in 
eq.(\ref{eq:Zchi}), since $\bar{u}(p)\gamma_5 u(p)=0$.

Now we turn to the wave function renormalizations
relevant to the  process under consideration.
We start with the vertices $Z\tilde\chi^+\tilde\chi^-$,
as shown in Fig.~\ref{GZchacha1lvertex}. If the one--loop graph is
attached to $\tilde\chi^+_b$ (Fig.~\ref{GZchacha1lvertex}d), 
we need to calculate the following amplitude:
\begin{eqnarray}
\overline{u}(k_2)[{\cal G}_{Z\chi\chi}^{ab}]^{2d}\,v(k_1)&=&
-i{g\over{2c_W}}\overline{u}(k_2)\Big[\widetilde\Sigma_{\chi\chi,1}^{bb}(k_2)
+\gamma_5\widetilde\Sigma_{\chi\chi,5}^{bb}(k_2) -\Delta m_{\chi_b} \Big]
\label{eq:WFRbZdiag}\\
&&\times{1\over{k_2^{\nu}\gamma_{\nu}-m_{\chi_b}}}\gamma^{\mu}
\Big[O'^L_{ab}(1-\gamma_5)+O'^R_{ab}(1+\gamma_5)\Big]v(k_1)
\nonumber
\end{eqnarray}
The term proportional to $\widetilde\Sigma_{\chi\chi,5}^{bb}(k_2)$ is 
simply evaluated by taking the spinors on shell and using
$\overline{u}(k_2)\gamma_5(k_2^{\nu}\gamma_{\nu}-m_{\chi_b})^{-1}=
\overline{u}(k_2)\gamma_5(-2m_{\chi_b})^{-1}$. For the term proportional to
$\widetilde\Sigma_{\chi\chi,1}^{bb}(k_2)$, more care is needed because of the
pole from the propagator acting on an on-shell spinor. The mass difference
$\Delta m_{\chi_b}$ appears because in the tree level amplitude a
tree level external chargino propagator is truncated by an on--shell
inverse propagator
\begin{equation}
{1\over{k_2^{\nu}\gamma_{\nu}-m_{\chi_b}(Q)}}
(k_2^{\nu}\gamma_{\nu}-m_{\chi_b})=
1+{{\Delta m_{\chi_b}}\over{k_2^{\nu}\gamma_{\nu}-m_{\chi_b}}}
+{\cal O}(2)
\label{eq:Deltam}
\end{equation}
introducing precisely the $\Delta m_{\chi_b}$ in eq.~(\ref{eq:WFRbZdiag})
and neglecting terms of two-loop order.
Of course, $\Delta m_{\chi_b}$ vanishes if we work in an on--shell scheme 
instead. Using eq.~(\ref{eq:ChaPoleRunM}) we see that this term gives rise 
to a factor of $\widetilde Z_{\chi_b}-1$, with $Z_{\chi_i}$ given in 
eq.~(\ref{eq:Zchi}). Combining this with the factor of 
$(\widetilde Z_{\chi_b})^{-1/2}$ for the external $\tilde\chi_b$ line 
we obtain the following contributions to the $F_{Z0}^{\pm}$ form factors:
\footnote{
The practical upshot of the above procedure is 
that the $\widetilde\Sigma_{\chi\chi,5}^{bb}(k_2)$ part of
Eq.\ref{eq:WFRbZdiag} contributes directly
while the remaining parts give a contribution equivalent to a factor
of $(\widetilde Z_{\chi_b})^{1/2}$ times the tree-level
amplitude. This simple result can be understood immediately from
LSZ reduction formula which requires us to take the on-shell limit
of the full propagator for each external leg,
as we did in Eq.\ref{eq:PropRen}, and then truncate each leg factor
$\widetilde Z_{\chi_i}/(p_{\mu}\gamma^{\mu}-m_{\chi_i})$,
and replace it by a factor of ${\widetilde Z_{\chi_i}}^{1/2}$ times the
appropriate spinor wavefunction for each leg.
Since the axial parts of the self-energy do not contribute to the
propagator near the pole their effect appears as a separate contribution.}
\begin{eqnarray}
F_{Z0}^+&=&-{g\over{2c_W}}O'^L_{ab}\bigg[
{1\over{2m_{\chi_b}}}(\widetilde A_{bb}^--\widetilde A_{bb}^+)-
\widetilde B_{bb}^--m_{\chi_b}
(\widetilde A'^+_{bb}+\widetilde A'^-_{bb})-m_{\chi_b}^2(
\widetilde B'^+_{bb}+\widetilde B'^-_{bb})\bigg]
\nonumber\\\label{eq:bChiChiFZ}\\
F_{Z0}^-&=&-{g\over{2c_W}}O'^R_{ab}\bigg[{1\over{2m_{\chi_b}}}
(\widetilde A_{bb}^+-\widetilde A_{bb}^-)-\widetilde B_{bb}^+-m_{\chi_b}
(\widetilde A'^+_{bb}+\widetilde A'^-_{bb})-m_{\chi_b}^2(
\widetilde B'^+_{bb}+\widetilde B'^-_{bb})\bigg]
\nonumber
\end{eqnarray}
where the prime represent the derivative with respect to $p^2$ and all
the functions are evaluated at $p^2=m_{\chi_b}^2$. In a similar way,
if the one--loop graph is attached to 
$\tilde\chi^+_a$ (Fig.~\ref{GZchacha1lvertex}e), the amplitude to be 
calculated is
\begin{eqnarray}
\overline{u}(k_2)[{\cal G}_{Z\chi\chi}^{ab}]^{2e}\,v(k_1)&=&
i{g\over{2c_W}}\overline{u}(k_2)\gamma^{\mu}\Big[O'^L_{ba}(1-\gamma_5)+
O'^R_{ba}(1+\gamma_5)\Big]{1\over{k_1^{\nu}\gamma_{\nu}+m_{\chi_a}}}
\nonumber\\
&&\times
\Big[\widetilde\Sigma_{\chi\chi,1}^{aa}(-k_1)+
\gamma_5\widetilde\Sigma_{\chi\chi,5}^{aa}(-k_1)
-\Delta m_{\chi_a}\Big]v(k_1)
\label{eq:WFRaZdiag}
\end{eqnarray}
Following similar steps to those for the self-energy insertion
on the $\tilde\chi_b^+$ line we find that 
these contributions to the $F_{Z0}^{\pm}$ form factors are
\begin{eqnarray}
F_{Z0}^+&=&-{g\over{2c_W}}O'^L_{ba}\bigg[
{1\over{2m_{\chi_a}}}(\widetilde A_{aa}^+-\widetilde A_{aa}^-)-
\widetilde B_{aa}^--m_{\chi_a}
(\widetilde A'^+_{aa}+\widetilde A'^-_{aa})-m_{\chi_a}^2(
\widetilde B'^+_{aa}+\widetilde B'^-_{aa})\bigg]
\nonumber\\\label{eq:aChiChiFZ}\\
F_{Z0}^-&=&-{g\over{2c_W}}O'^R_{ba}\bigg[{1\over{2m_{\chi_a}}}
(\widetilde A_{aa}^--\widetilde A_{aa}^+)-\widetilde B_{aa}^+-m_{\chi_a}
(\widetilde A'^+_{aa}+\widetilde A'^-_{aa})-m_{\chi_a}^2(
\widetilde B'^+_{aa}+\widetilde B'^-_{aa})\bigg]
\nonumber
\end{eqnarray}
and every function is evaluated at $m_{\chi_a}^2$.

We now turn to the vertices $\gamma\chi^+\chi^-$. The procedure is 
analogous and we just give the final results. If the one--loop
graph is attached to the $\tilde\chi^+_b$ the we get the following
contributions to the $F_{\gamma0}$ form factors
\begin{eqnarray}
F_{\gamma 0}^+ &=&{e\over 2}\bigg[
{1\over{2m_{\chi_b}}}(\widetilde A_{bb}^--\widetilde A_{bb}^+)-
\widetilde B_{bb}^--m_{\chi_b}
(\widetilde A'^+_{bb}+\widetilde A'^-_{bb})-m_{\chi_b}^2(
\widetilde B'^+_{bb}+\widetilde B'^-_{bb})\bigg]\delta_{ab}
\nonumber\\\label{eq:bChiChiFG}\\
F_{\gamma 0}^- &=&{e\over 2}\bigg[{1\over{2m_{\chi_b}}}
(\widetilde A_{bb}^+-\widetilde A_{bb}^-)-\widetilde B_{bb}^+-m_{\chi_b}
(\widetilde A'^+_{bb}+\widetilde A'^-_{bb})-m_{\chi_b}^2(
\widetilde B'^+_{bb}+\widetilde B'^-_{bb})\bigg]\delta_{ab}
\nonumber
\end{eqnarray}
where every function is evaluated at $m_{\chi_b}^2$. Now, if the one--loop
graph is attached to $\tilde\chi^+_a$ we get
\begin{eqnarray}
F_{\gamma 0}^+ &=&{e\over 2}\bigg[
{1\over{2m_{\chi_a}}}(\widetilde A_{aa}^--\widetilde A_{aa}^+)-
\widetilde B_{aa}^+-m_{\chi_a}
(\widetilde A'^+_{aa}+\widetilde A'^-_{aa})-m_{\chi_a}^2(
\widetilde B'^+_{aa}+\widetilde B'^-_{aa})\bigg]\delta_{ab}
\nonumber\\\label{eq:aChiChiFG}\\
F_{\gamma 0}^- &=&{e\over 2}\bigg[{1\over{2m_{\chi_a}}}
(\widetilde A_{aa}^+-\widetilde A_{aa}^-)-\widetilde B_{aa}^--m_{\chi_a}
(\widetilde A'^+_{aa}+\widetilde A'^-_{aa})-m_{\chi_a}^2(
\widetilde B'^+_{aa}+\widetilde B'^-_{aa})\bigg]\delta_{ab}
\nonumber
\end{eqnarray}
where every function is evaluated at $m_{\chi_a}^2$.

In the case of $e^{\pm}\tilde\nu_e\tilde\chi^{\mp}_b$ vertices, the
procedure is analogous, with the only extra complication given by the 
handling of the charge conjugation matrix $C$ in the Feynman rules. If
the one--loop graph is attached to the $\tilde\chi^+_b$ the we get
the following contribution to $F_{\tilde\nu_e}^+$
\begin{equation}
F_{\tilde\nu_e}^+ ={g\over 2}V_{b1}\bigg[
{1\over{2m_{\chi_b}}}(\widetilde A_{bb}^--\widetilde A_{bb}^+)-
\widetilde B_{bb}^--m_{\chi_b}
(\widetilde A'^+_{bb}+\widetilde A'^-_{bb})-m_{\chi_b}^2(
\widetilde B'^+_{bb}+\widetilde B'^-_{bb})\bigg]
\label{eq:bChiChiFS}            
\end{equation}
with every function is evaluated at $m_{\chi_b}^2$. And finally, if
the one--loop graph is attached to $\tilde\chi^+_a$ we get a
contribution to $F_{\tilde\nu_e}^-$ given by
\begin{equation}
F_{\tilde\nu_e}^- =-{g\over 2}V_{a1}^*\bigg[
{1\over{2m_{\chi_a}}}(\widetilde A_{aa}^+-\widetilde A_{aa}^-)-
\widetilde B_{aa}^+-m_{\chi_a}
(\widetilde A'^+_{aa}+\widetilde A'^-_{aa})-m_{\chi_a}^2(
\widetilde B'^+_{aa}+\widetilde B'^-_{aa})\bigg]
\label{eq:aChiChiFS}
\end{equation}
with every function is evaluated at $m_{\chi_a}^2$.

\newpage

\section{Results}

\begin{figure}
\centerline{\protect\hbox{\psfig{file=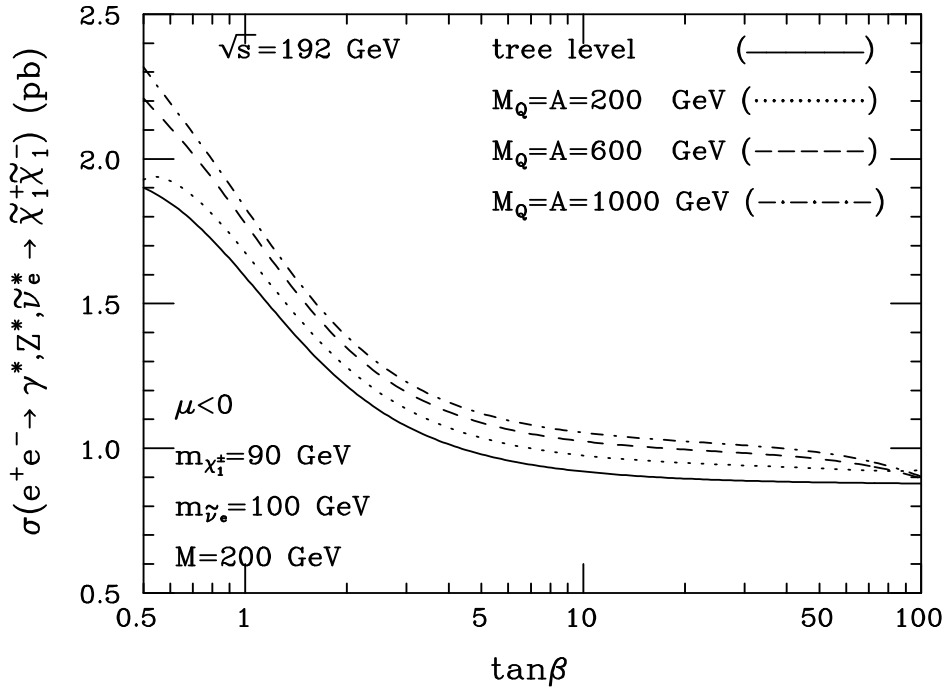,height=11cm,width=1.0\textwidth,angle=90}}}
\caption{One--loop and tree level chargino production cross section as a
function of $\tan\beta$, for 192 GeV center of mass energy.} 
\label{fig:c192tb} 
\end{figure}

\begin{figure}
\centerline{\protect\hbox{\psfig{file=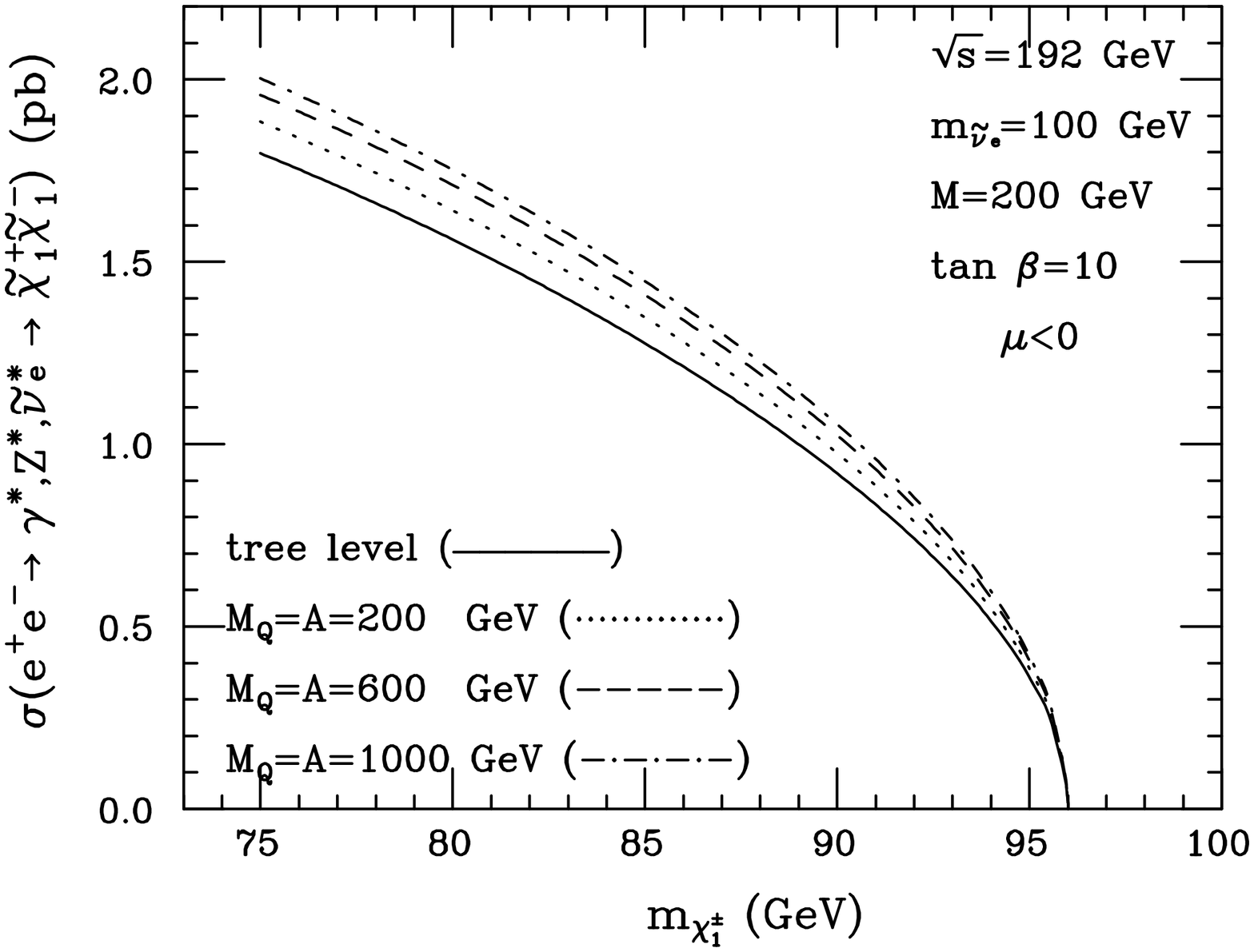,height=11cm,width=1.0\textwidth,angle=90}}}
\caption{One--loop and tree level chargino production cross section as a
function of the chargino mass $m_{\chi^{\pm}_1}$, for 192 GeV center of mass 
energy.} 
\label{fig:c192mc1}
\end{figure}

\begin{figure}
\centerline{\protect\hbox{\psfig{file=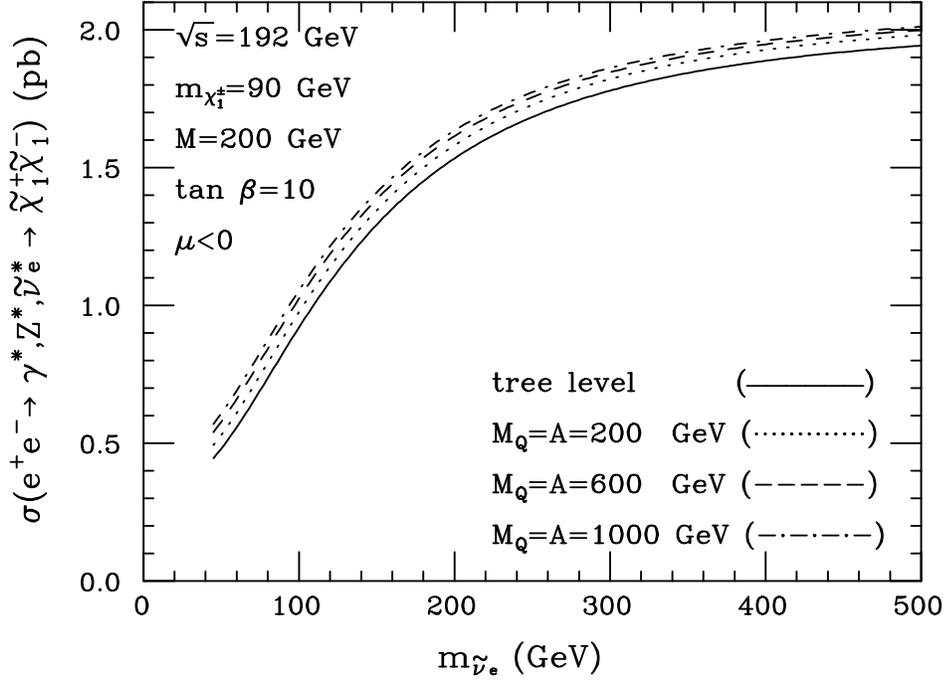,height=11cm,width=1.0\textwidth,angle=90}}}
\caption{One--loop and tree level chargino production cross section as a
function of the sneutrino mass $m_{\tilde\nu_e}$, for 
192 GeV center of mass energy.} 
\label{fig:c192msn}
\end{figure}

\begin{figure}
\centerline{\protect\hbox{\psfig{file=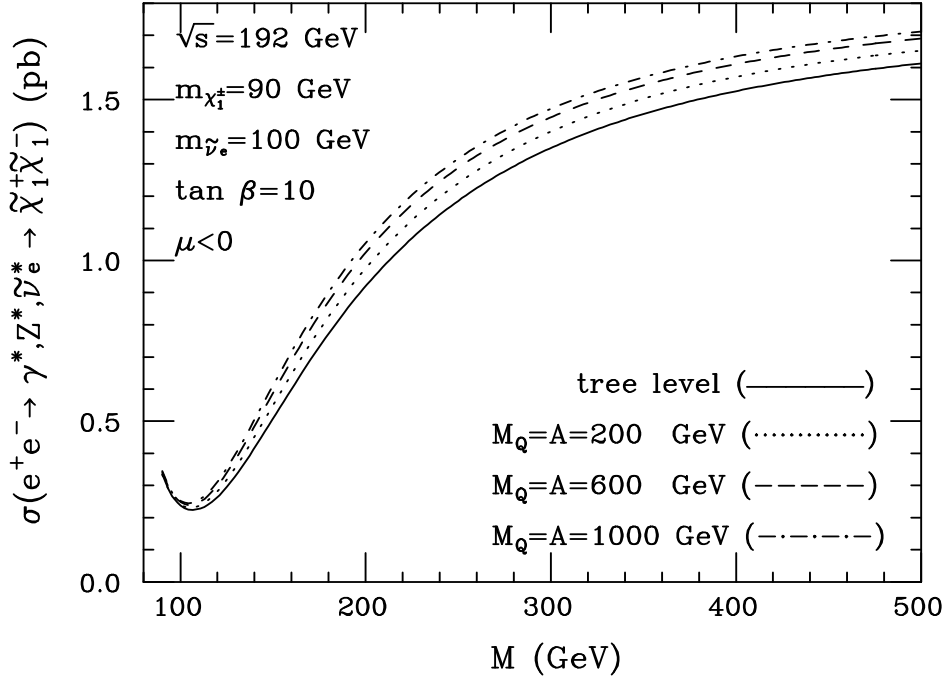,height=11cm,width=1.0\textwidth,angle=90}}}
\caption{One--loop and tree level chargino production cross section as a
function of the $SU(2)$ gaugino mass $M$, for 192 GeV center of mass energy.} 
\label{fig:c192m}
\end{figure}

In this section we calculate numerically the radiatively corrected 
chargino production cross section and compare it with the tree level.
We start with a center of mass energy of $\sqrt{s}=192$ GeV relevant
for LEP2, and consider the case $\mu<0$, where $\mu$ is the supersymmetric
Higgs mass parameter in the superpotential. Figs.~\ref{fig:c192tb} to
\ref{fig:c192m} correspond to this center of mass energy.
Radiative corrections to this 
cross section are parametrized by the squark soft masses which we take 
degenerate $M_Q=M_U=M_D$, and by the trilinear soft mass parameters 
$A\equiv A_U=A_D$, also taken degenerate. This choice is done at the 
weak scale and it is made for simplicity, \ie, should not be confused
with universality of minimal supergravity at the unification scale.

In Fig.~\ref{fig:c192tb} we plot 
$\sigma (e^+e^-\rightarrow \tilde{\chi}^+_1 \tilde{\chi}^-_1)$ as a function
of $\tan\beta$, for a constant value of the chargino mass 
$m_{\chi^{\pm}_1}=90$ GeV, the sneutrino mass $m_{\tilde\nu_e}=100$ GeV,
and the gaugino mass $M=200$ GeV. The tree level cross section is in
the solid line and decreases from 1.9 pb. to 0.9 pb. when $\tan\beta$ 
increases from 0.5 to 100. Three radiatively corrected curves are presented,
and they are parametrized by $M_Q=A=200$ GeV (dots), $M_Q=A=600$ GeV 
(dashes), and $M_Q=A=1$ TeV (dotdashes). We observe that radiative 
corrections are positive and grow with the squark mass parameters. The
maximum value of the corrections vary from $5\%$ at high $\tan\beta$ to
$22\%$ at small $\tan\beta$. A logarithmic growth of quantum corrections
with the squark mass parameters is observed, as it should be. Nevertheless,
this logarithmic dependence is lost if $\tan\beta<1$ or $\tan\beta\gsim 30$,
as can be seen from the figure. The reason is that at those values
of $\tan\beta$ there is no longer a single scale of squark masses,
specially for $M_Q=A=200$ GeV where the squark masses range from 100 GeV
to 370 GeV. It is worth pointing out that the value of $\mu$ is not 
constant along the curves because it is fixed by the constant value of the 
chargino mass $m_{\tilde\chi^{\pm}_1}$.

In Fig.~\ref{fig:c192mc1} we explore the dependence of the radiatively 
corrected cross section 
$\sigma (e^+e^-\rightarrow \tilde{\chi}^+_1 \tilde{\chi}^-_1)$ on 
the chargino mass $m_{\chi^{\pm}_1}$ for $\sqrt{s}=192$ GeV. We keep 
constant the sneutrino mass $m_{\tilde\nu_e}=100$ GeV, the gaugino mass
$M=200$ GeV, and $\tan\beta=10$. The tree level cross section decreases
from 1.8 pb. for $m_{\chi^{\pm}_1}=75$ GeV to 0.33 pb. for 
$m_{\chi^{\pm}_1}=95$ GeV and continues to zero at the kinematic limit.
Quantum corrections, on the other hand, grow with the 
chargino mass from $11\%$ for $m_{\chi^{\pm}_1}=75$ GeV to $16\%$ for 
$m_{\chi^{\pm}_1}=95$ GeV when $M_Q=A=1$ TeV. The chargino mass 
$m_{\chi^{\pm}_1}$ in this section corresponds to the pole mass given
in eq.~(\ref{eq:ChaPoleRunM}). In Fig.~\ref{fig:c192mc1}, the running
chargino mass is $1\%$ smaller than the pole mass for $M_Q=A=1$ TeV and
the correction decreases with smaller squark mass parameters.

The dependence of the cross section 
$\sigma (e^+e^-\rightarrow \tilde{\chi}^+_1 \tilde{\chi}^-_1)$ on the
electron--type sneutrino mass is shown in Fig.~\ref{fig:c192msn}, where we 
take the chargino mass $m_{\chi^{\pm}_1}=90$ GeV, the gaugino mass 
$M=200$ GeV, and $\tan\beta=10$. Charginos can be 
produced in the t--channel with an intermediate $\tilde\nu_e$, and in
the s--channel with intermediate $Z$ and $\gamma$. The interference
between the t--channel and the s--channel is negative, and this is the 
reason why the cross section decreases when the sneutrino becomes lighter.
The tree level cross section varies from 0.45 pb. when 
$m_{\tilde\nu_e}=45$ GeV to 1.9 pb when $m_{\tilde\nu_e}=500$ GeV.
On the other hand, radiative corrections are larger when 
the sneutrino is lighter, going from $28\%$ for $m_{\tilde\nu_e}=45$ GeV
to $3.5\%$ for $m_{\tilde\nu_e}=500$ GeV if $M_Q=A=1$ TeV. Smaller
corrections are found if the squark mass parameters are decreased.

The last graph with center of mass energy $\sqrt{s}=192$ GeV is 
Fig.~\ref{fig:c192m}, where we plot the chargino production cross section
$\sigma (e^+e^-\rightarrow \tilde{\chi}^+_1 \tilde{\chi}^-_1)$
as a function of the $SU(2)$ gaugino mass $M$, while keeping constant the
chargino mass $m_{\chi^{\pm}_1}=90$ GeV, the sneutrino mass 
$m_{\tilde\nu_e}=100$ GeV, and $\tan\beta=10$. 
The tree level cross section
decreases from 1.6 pb. when $M=500$ GeV to a minimum of 0.22 pb. at
around $M=105$ GeV, and grows again up to 0.34 pb. at $M=90$ GeV.
Below this value of the gaugino mass $M$ there is no solution for
$\mu<0$ which gives $m_{\chi^{\pm}_1}=90$ GeV. As before, the largest 
quantum corrections are found with $M_Q=A=1$ TeV. For $M$ close to
90 GeV the corrections are only of a few percent, but they grow fast
until a maximum of $21\%$ at $M=140$ GeV. For larger values of the gaugino 
mass, the corrections slowly decrease until they reach the value $6\%$
at $M=500$ GeV.

In summary we can say that for a center of mass energy $\sqrt{s}=192$ GeV,
relevant for LEP2, the radiative corrections to the production cross section
of a pair of light charginos grow logarithmically with the squark mass
parameters. The corrections are positive and typically $10\%$ to $15\%$ for
$M_Q=A=1$ TeV, $8\%$ to $11\%$ for $M_Q=A=600$ GeV, and $3\%$ to $5\%$ for
$M_Q=A=200$ GeV, with a maximum value of the order of $30\%$,  $20\%$,
and $10\%$ respectively, for the region of parameter space explored here.  

\begin{figure}
\centerline{\protect\hbox{\psfig{file=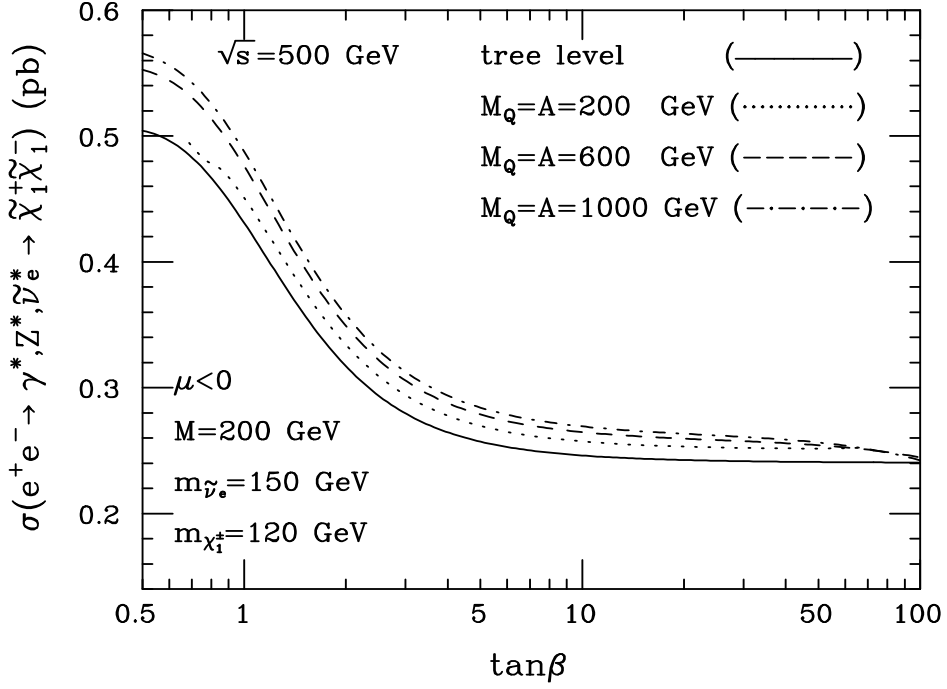,height=11cm,width=1.0\textwidth,angle=90}}}
\caption{One--loop and tree level chargino production cross section as a
function of $\tan\beta$, for 500 GeV center of mass energy.} 
\label{fig:c5tb}
\end{figure}
\begin{figure}
\centerline{\protect\hbox{\psfig{file=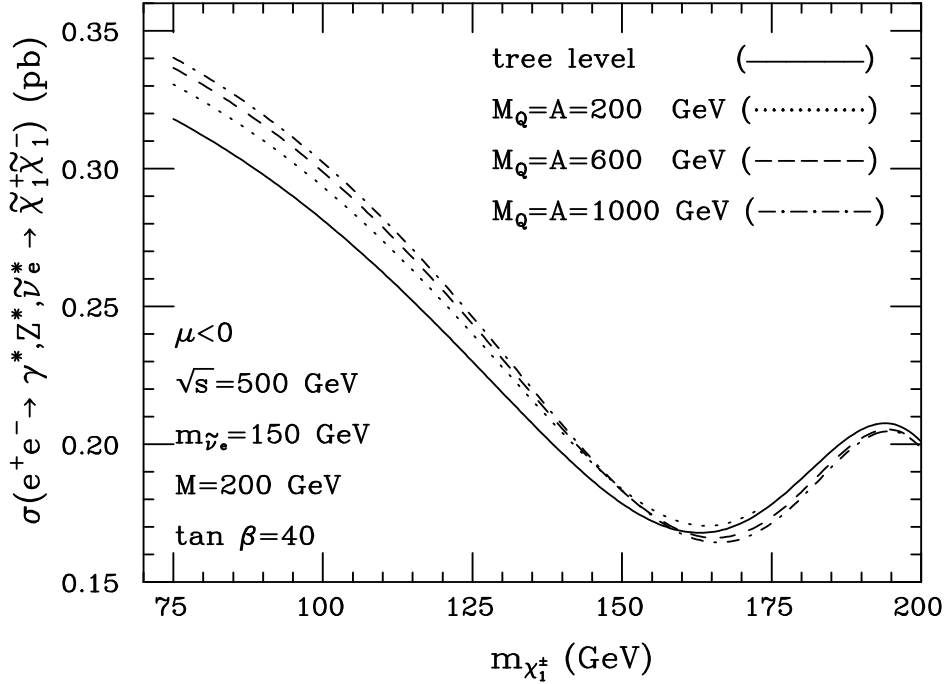,height=11cm,width=1.0\textwidth,angle=90}}}
\caption{One--loop and tree level chargino production cross section as a
function of of the chargino mass $m_{\chi^{\pm}_1}$, for 500 GeV center of 
mass energy.} 
\label{fig:c5mc1}
\end{figure}

\begin{figure}
\centerline{\protect\hbox{\psfig{file=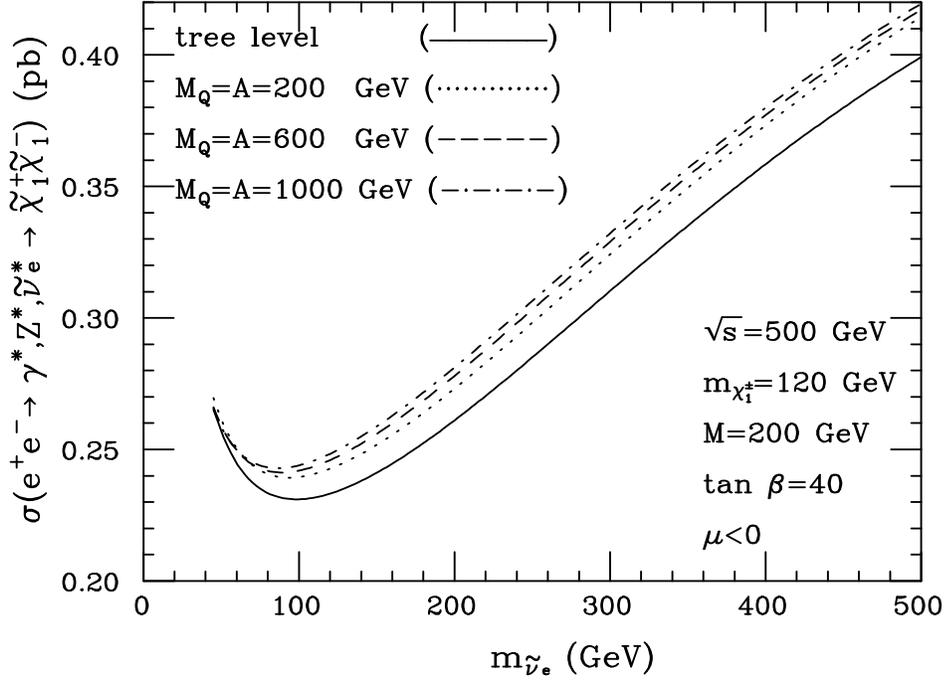,height=11cm,width=1.0\textwidth,angle=90}}}
\caption{One--loop and tree level chargino production cross section as a
function of the sneutrino mass $m_{\tilde\nu_e}$, for 500 GeV 
center of mass energy.} 
\label{fig:c5msn}
\end{figure}

\begin{figure}
\centerline{\protect\hbox{\psfig{file=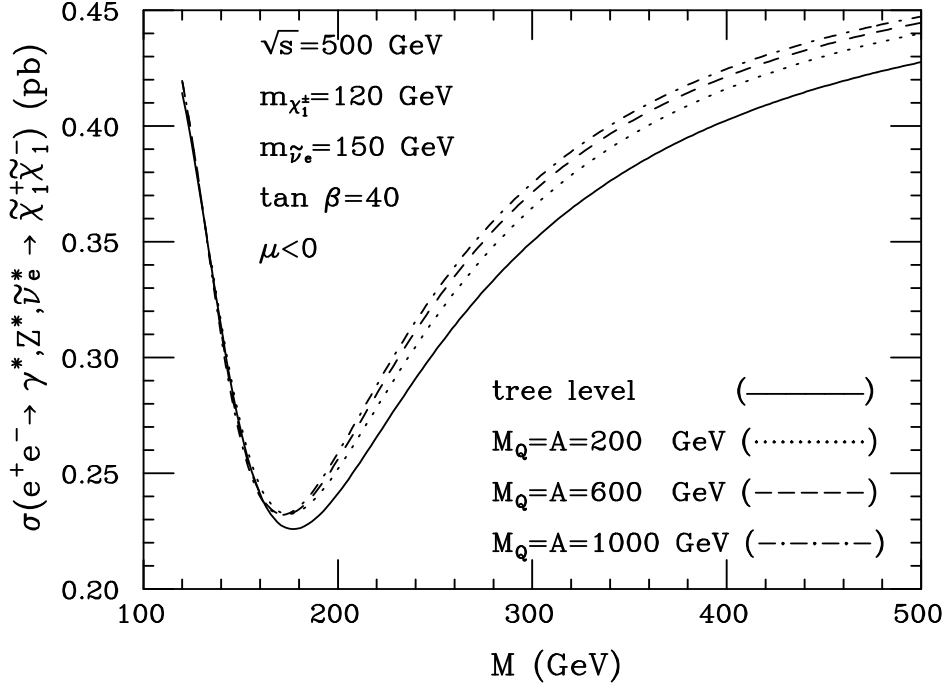,height=11cm,width=1.0\textwidth,angle=90}}}
\caption{One--loop and tree level chargino production cross section as a
function of the $SU(2)$ gaugino mass $M$, for 500 GeV 
center of mass energy.} 
\label{fig:c5m}
\end{figure}

 
Now we turn to a center of mass energy given by $\sqrt{s}=500$ GeV, relevant 
for future $e^+e^-$ colliders. Fig.~\ref{fig:c5tb} to \ref{fig:c5m}
correspond to this energy, where we continue working with $\mu<0$.

In Fig.~\ref{fig:c5tb} we plot the total cross
section $\sigma(e^+e^-\rightarrow\tilde{\chi}^+_1\tilde{\chi}^-_1)$
as a function of $\tan\beta$ for a fixed value of the chargino mass
$m_{\chi^{\pm}_1}=120$ GeV, the sneutrino mass $m_{\tilde\nu_e}=150$ GeV, 
and the gaugino mass $M=200$ GeV. 
The behaviour of the tree--level cross section is similar to the same curve
with the previous center of mass energy, decreasing from 0.5 pb. at
low $\tan\beta$ to 0.24 pb. at large $\tan\beta$. The corrections are
positive and grow logarithmically with the squark mass parameters in
the central region of $\tan\beta$, but this is not so in the extremes
for the reason explained before. Furthermore, the curve corresponding to
$M_Q=A=200$ GeV has to be truncated because if $\tan\beta\lsim 0.7$ we
have $m_{\tilde t_1}\lsim 100$ GeV, and if $\tan\beta\gsim 70$ we get
$m_{\tilde b_1}\lsim 100$ GeV. The maximum value of the corrections
occurs at a point slightly over $\tan\beta=1$ and corresponds to $13\%$,
$10\%$, and $5\%$ for squark mass parameters equal to 1000, 600, and 200
GeV respectively.

The dependence of the total cross section 
$\sigma(e^+e^-\rightarrow\tilde{\chi}^+_1\tilde{\chi}^-_1)$ as a 
function of the chargino mass is shown in Fig.~\ref{fig:c5mc1}. We consider 
the case $m_{\tilde\nu_e}=150$ GeV, $M=200$ GeV, and $\tan\beta=40$. The
values of the chargino mass shown are the physical pole mass, which 
is at most $1.6\%$ larger than the running mass. Values of 
$m_{\tilde\chi_1^{\pm}}$ larger than 200 GeV are not displayed because
there is no solution for $\mu$ compatible with those values of the 
chargino mass. In fact, $|\mu|$ grows as we approach to 
$m_{\tilde\chi_1^{\pm}}=200$ GeV and diverges at that point. Furthermore,
large values of $|\mu|$ and the large value of $\tan\beta$ chosen 
induce a large sbottom mixing, and as a consequence the curve corresponding 
to $M_Q=A=200$ GeV is truncated because for example we get 
$m_{\tilde b_1}\lsim 100$ if $m_{\tilde\chi_1^{\pm}}\gsim 174$ GeV.
Radiative corrections for $M_Q=A=1$ TeV are positive for small chargino
masses and reach a maximum value of $8\%$, and they are negative for
large $m_{\tilde\chi_1^{\pm}}$ with a maximum value of $-4\%$.

In Fig.~\ref{fig:c5msn} we plot 
$\sigma(e^+e^-\rightarrow\tilde{\chi}^+_1\tilde{\chi}^-_1)$ in terms
of the sneutrino mass $m_{\tilde\nu_e}$, for a constant value of 
the chargino mass $m_{\tilde\chi_1^{\pm}}=120$ GeV, the gaugino mass
parameter $M=200$ GeV, and $\tan\beta=40$. 
Due to the negative interference 
between the s--channel and the t--channel chargino production, the
total cross section has a minimum at a certain value of the sneutrino mass.
This minimum is shifted by radiative corrections from 
$m_{\tilde\nu_e}\approx 100$ GeV at tree level to 
$m_{\tilde\nu_e}\approx 90$ GeV at one--loop. Quantum corrections are
larger for $M_Q=A=1$ TeV and have a maximum value of $8\%$ for medium
values of the sneutrino mass.

In the last graph corresponding to $\sqrt{s}=500$ GeV we have the 
production cross section 
$\sigma(e^+e^-\rightarrow\tilde{\chi}^+_1\tilde{\chi}^-_1)$ as a function
of the gaugino mass $M$, and it is plotted in Fig.~\ref{fig:c5m}. 
We consider $m_{\tilde\chi_1^{\pm}}=120$ GeV, $m_{\tilde\nu_e}=150$
GeV, and $\tan\beta=40$. The tree level cross section has a minimum
at $M\approx 180$ GeV which is displaced to $M\approx 173$ GeV by
radiative corrections. These corrections are positive most of the time,
reaching a maximum of $9\%$ when $M_Q=A=1$ TeV, but can be slightly 
negative at small values of the gaugino mass. The value of $|\mu|$
grows as we approach $M=120$ GeV, and consequently the curve corresponding
to $M_Q=A=200$ GeV is truncated because $m_{\tilde t_1}$ is too light.

Now we summarize the results corresponding to a center of mass energy
given by $\sqrt{s}=500$ GeV. As before, for the region of parameter 
space explored here, we observe a logarithmic growth of the radiative
corrections with the squark mass parameters only for medium values
of $\tan\beta$. For extreme values of $\tan\beta$ there is no longer a
unique squark mass scale due to large mass splitings. Corrections are 
smaller than the ones found for the previous center of mass energy, 
reaching maximum values of $13\%$, $11\%$, and $5\%$ for squark mass 
parameters equal to 1000, 600, and 200 GeV respectively. Contrary to the
previous case, we find here negative corrections, reaching extreme values
of $-4\%$, $-2.5\%$, and $-0.2\%$ respectively.

\begin{figure}
\centerline{\protect\hbox{\psfig{file=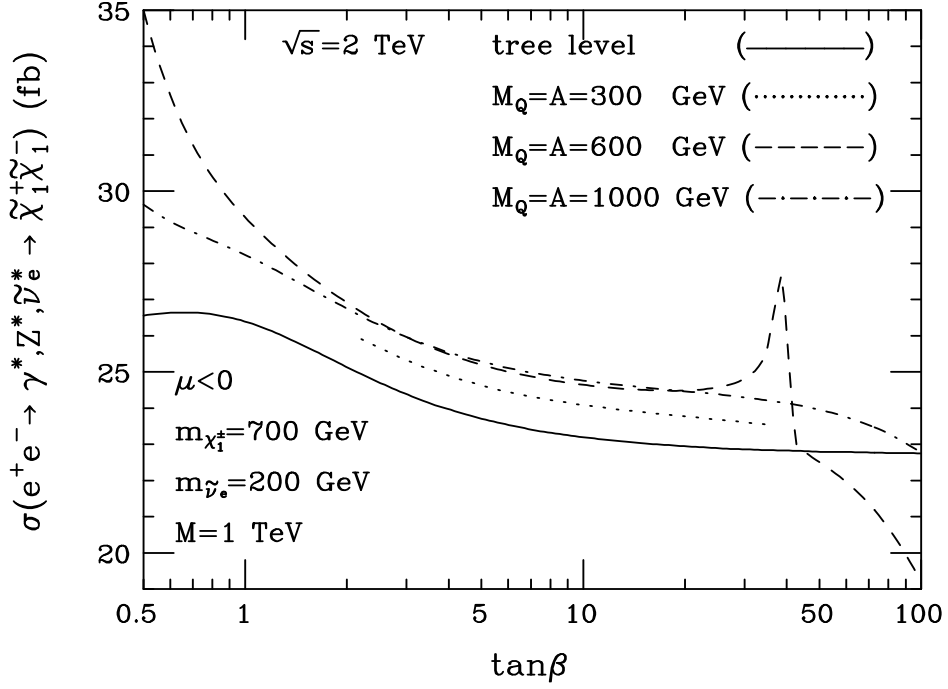,height=11cm,width=1.0\textwidth,angle=90}}}
\caption{One--loop and tree level chargino production cross section as a
function of $\tan\beta$, for 2 TeV center of mass energy.} 
\label{fig:c2tb}
\end{figure}

\begin{figure}
\centerline{\protect\hbox{\psfig{file=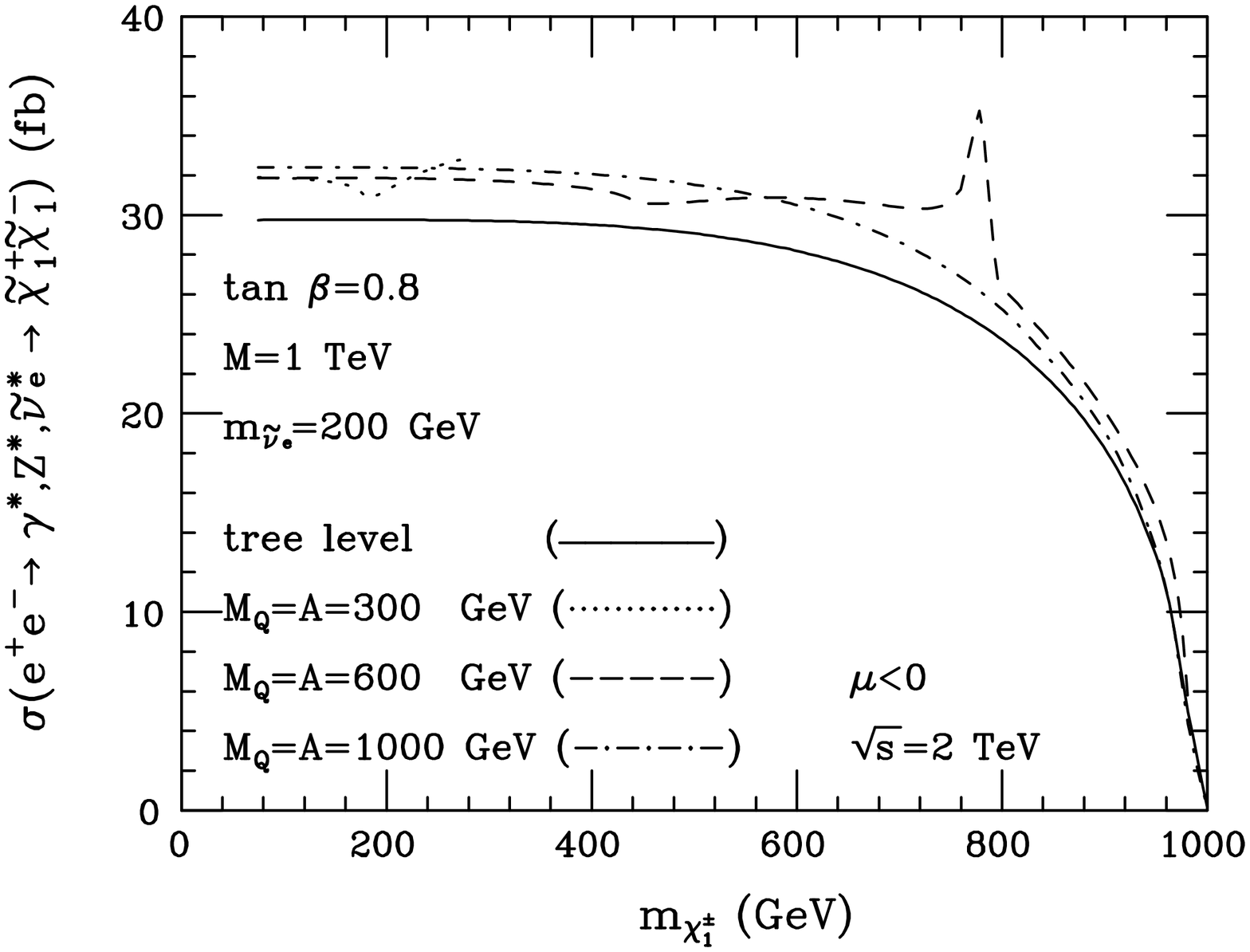,height=11cm,width=1.0\textwidth,angle=90}}}
\caption{One--loop and tree level chargino production cross section as a
function of of the chargino mass $m_{\chi^{\pm}_1}$, for 2 TeV center of 
mass energy.} 
\label{fig:c2mc1}
\end{figure}

\begin{figure}
\centerline{\protect\hbox{\psfig{file=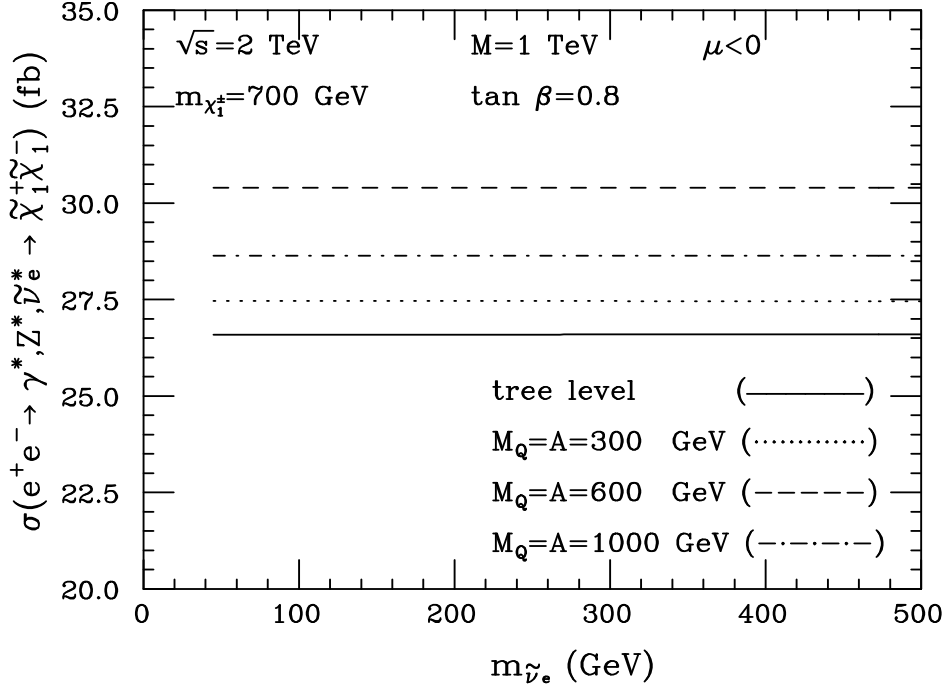,height=11cm,width=1.0\textwidth,angle=90}}}
\caption{One--loop and tree level chargino production cross section as a
function of the sneutrino mass $m_{\tilde\nu_e}$, for 2 TeV 
center of mass energy.} 
\label{fig:c2msn}
\end{figure}

\begin{figure}
\centerline{\protect\hbox{\psfig{file=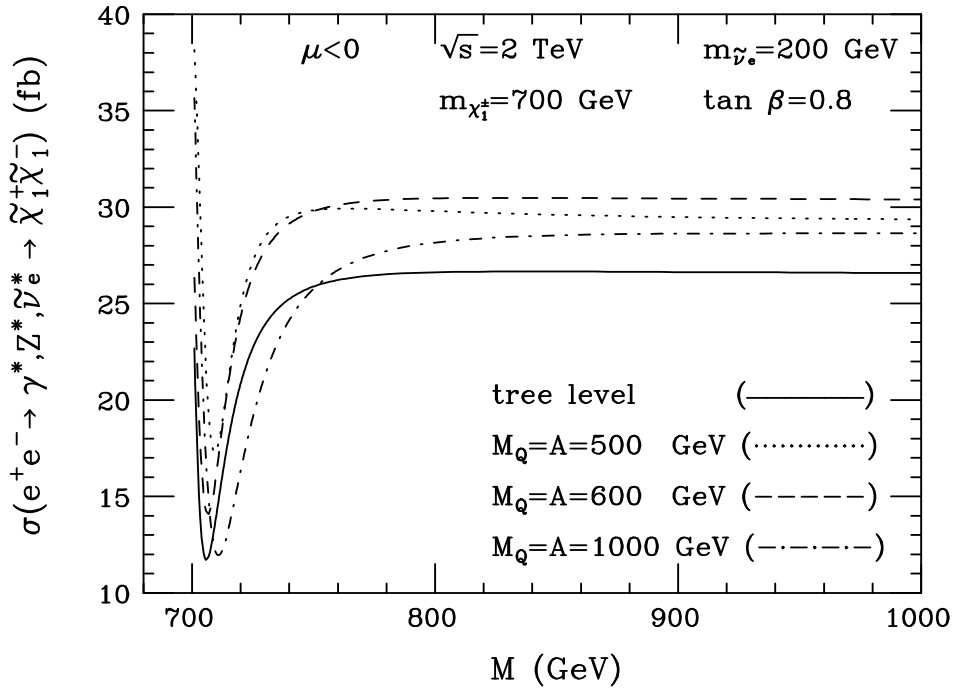,height=11cm,width=1.0\textwidth,angle=90}}}
\caption{One--loop and tree level chargino production cross section as a
function of the $SU(2)$ gaugino mass $M$, for 2 TeV center of mass energy.} 
\label{fig:c2m}
\end{figure}

The last value of the center of mass energy we analyze here corresponds
to an electron--positron collider with $\sqrt{s}=2$ TeV. Relevant
to this energy we have Figs.~\ref{fig:c2tb} to \ref{fig:c2m}. Again, we 
study the production of a pair of light charginos with $\mu<0$, but
this time we concentrate on a very heavy chargino with 
$m_{\tilde\chi_1^{\pm}}=700$ GeV.

We plot in Fig.~\ref{fig:c2tb} the total cross section 
$\sigma(e^+e^-\rightarrow\tilde{\chi}^+_1\tilde{\chi}^-_1)$ in
terms of $\tan\beta$ for $\sqrt{s}=2$ TeV. We take 
$m_{\tilde\chi_1^{\pm}}=700$ GeV, $m_{\tilde\nu_e}=200$, and $M=1$ TeV.
The lowest value of squark mass parameters we consider is $M_Q=A=300$ GeV
(dotted line). This is done in order to have reasonable squark masses.
In fact, acceptable squark masses are found only if 
$2\lsim\tan\beta\lsim 35$ and the restrictions on $\tan\beta$ are 
stronger for smaller squark mass parameters.
The peak observed in the curve $M_Q=A=600$ GeV at $\tan\beta\approx 40$
is physical and corresponds to the threshold 
$m_{\tilde\chi_1^{\pm}}=m_t+m_{\tilde b_1}$ where a chargino can decay
on--shell to a top quark plus the lightest bottom squark $\tilde b_1$.
The largest radiative corrections are found for $M_Q=A=600$ GeV and can
have either sign. The extreme values are $32\%$ at low $\tan\beta$ and
$-16\%$ at high $\tan\beta$.

In Fig.~\ref{fig:c2mc1} we have
$\sigma(e^+e^-\rightarrow\tilde{\chi}^+_1\tilde{\chi}^-_1)$ as a 
function of the pole chargino mass $m_{\tilde\chi_1^{\pm}}$ for a constant
value of the sneutrino mass $m_{\tilde\nu_e}=200$, the gaugino mass 
$M=1$ TeV, and $\tan\beta=0.8$. The pole chargino mass is at most $1.5\%$
smaller that the running mass. The curve corresponding to 
$M_Q=A=300$ GeV is truncated because for $m_{\tilde\chi_1^{\pm}}\gsim 260$
GeV we have $m_{\tilde t_1}\lsim 100$ GeV.
The peak in the curve $M_Q=A=600$ GeV close to $m_{\tilde\chi_1^{\pm}}=780$
GeV is the same threshold found in the previous figure where
$m_{\tilde\chi_1^{\pm}}=m_t+m_{\tilde b_1}$. Radiative corrections are larger
for $M_Q=A=600$ GeV and can go up to $44\%$, but with typical values
of the order of $10\%$. They can also be negative reaching values of 
$-12\%$ for $M_Q=A=1$ TeV.

The dependence of the total cross section 
$\sigma(e^+e^-\rightarrow\tilde{\chi}^+_1\tilde{\chi}^-_1)$ on the
sneutrino mass $m_{\tilde\nu_e}$ is shown in Fig.~\ref{fig:c2msn}, where we
keep constant the value of the chargino mass $m_{\tilde\chi_1^{\pm}}=700$ 
GeV, the $SU(2)$ gaugino mass $M=1$ TeV, and $\tan\beta=0.8$. We 
observe that the cross section depends very weakly on the sneutrino mass.
The reason is that the sneutrino contribution is negligible because
the lightest chargino is almost purely higgsino and therefore its
coupling to an electron plus a sneutrino is very small. Since
the $W$--boson mass is so small compared with $M$ and $\mu$, the
sneutrino contribution would be sizable only if we take a gaugino mass
close to the chargino mass. Quantum corrections in this case are positive,
almost constant, and equal to $8\%$, $14\%$, and $3\%$ if the squark
mass parameters are equal to 1000, 600, and 300 GeV respectively. The case
$M_Q=A=600$ GeV is larger due to the proximity of the 
$m_{\tilde\chi_1^{\pm}}=m_t+m_{\tilde b_1}$ threshold (see 
Fig.~\ref{fig:c2mc1}).

The final plot with $\sqrt{s}=2$ TeV is Fig.~\ref{fig:c2m}, where we 
have the total cross section
$\sigma(e^+e^-\rightarrow\tilde{\chi}^+_1\tilde{\chi}^-_1)$
as a function of the gaugino mass $M$. The cross section is fairly constant
for $M\gsim 750$ GeV and has a deep minimum centered at $M\approx 705$ GeV
at tree level, which is shifted upwards in a few GeV by quantum 
corrections. In the region of constant cross section, radiative corrections
are positive and larger for $M_Q=A=600$ GeV, and are of the order of 
$15\%$. In the region around the minimum, radiative corrections are even 
larger and can have either sign, with extreme values that can go up
to $100\%$ and down to $-26\%$.

We summarize now our findings at $\sqrt{s}=2$ TeV. In this case, due to the
fact that the chargino mass is heavy and can decay into on--shell squarks
plus quarks, we find large radiative corrections close to the threshold.
This makes the radiatively corrected cross section less regular than the 
previous cases, and with larger extreme values. We find that the 
maximum value for the correction can go up to $100\%$, and that can be 
negative also, with a extreme value of $-26\%$. Typically, we find 
corrections in the range between $-20\%$ to $20\%$. 

\section{Conclusions}

We have calculated the one--loop radiative corrections to the total
production cross section of a pair of charginos in the MSSM. Using the
diagramatic method we give formulas for the radiatively corrected
$\sigma(e^+e^-\rightarrow\tilde{\chi}^+_b\tilde{\chi}^-_a)$, where
$\tilde{\chi}^+_b$ and $\tilde{\chi}^-_a$ are any of the two charginos.
We work in the $\overline{MS}$ scheme and under the approximation 
where only top and bottom quarks and squarks are considered in the
loops. We organize the calculation introducing form factors which
renormalize the $Z\tilde\chi^+_b\tilde{\chi}^-_a$, 
$\gamma\tilde\chi^+_b\tilde{\chi}^-_a$, and 
$e^{\pm}\tilde\nu_e\tilde\chi^{\pm}_a$ vertices. Numerically, we concentrate
on the production of a pair of light charginos in electron--positron
colliders with center of mass energy given by 192, 500, and 2000 GeV
and with $\mu<0$. Radiative corrections to
$\sigma(e^+e^-\rightarrow\tilde{\chi}^+_1\tilde{\chi}^-_1)$ are
parametrized by the value of the squark mass parameters, which for 
simplicity we
take equal at the weak scale $M_Q=M_U=M_D$ and $A\equiv A_U=A_D$.
We explore the cross section varying the parameters which affect the
chargino cross section at tree level: the ratio of vacuum expectation values
$\tan\beta$, the $SU(2)$ gaugino mass $M$, the electron--type sneutrino
mass $m_{\tilde\nu_e}$, and the lightest chargino mass 
$m_{\tilde\chi_1^{\pm}}$, which has been used instead of the 
supersymmetric mass parameter $\mu$ as an independent variable. For a 
center of mass energy $\sqrt{s}=192$ GeV we have found positive radiative 
corrections which are typically of $10\%$ to $15\%$ if $M_Q=A=1$ TeV and 
with a maximum value of $30\%$. For a center of mass energy of 
$\sqrt{s}=500$ GeV we find corrections with a maximum value of $13\%$
if $M_Q=A=1$ TeV, but also negative corrections are found, and with
an extreme value of $-4\%$. If $\sqrt{s}=2$ TeV, radiative corrections
are larger, and extreme values of $100\%$ and $-26\%$ can be found.
No big differences are observed for the case $\mu>0$.

The calculation of the radiative corrections to the chargino pair 
production cross section reported here will enable us to interpret
more reliably the chargino searches performed al LEP2 as lower limits
on the chargino mass and, consequently, as exclusion zones in the
parameter space. If charginos are discovered, this calculation will be 
essential in order to extract the value of the fundamental parameters
of the theory from the experimental measurements.

\section*{Acknowledgements}

The authors are grateful to PPARC for partial support under contract 
no. GR/K55738.
M.A.D. was also supported by a DGICYT postdoctoral grant of the spanish
Ministerio de Educaci\'on y Ciencia.

\section*{Appendix A: Feynman Rules}

\setcounter{equation}{0}
\renewcommand{\theequation}{A.\arabic{equation}}

In this Appendix we list the Feynman rules for the MSSM that we 
require in our calculations. These rules are adapted from Haber and Kane
\cite{HaberKane} and Gunion and Haber \cite{GunHaber}, although we have 
introduced a more compact notation for the vertices involving mass 
eigenstates as we shall explain.

Apart from the Standard Model Feynman rules, which we do not
list here, there are three classes of vertex which are encountered
in our calculations:
\begin{enumerate}
\item{The $Z$ (and photon) - chargino - chargino  vertices 
%(Fig.~\ref{GZchachaFeyn})
}
\item{The $Z$ (and photon) - squark - squark  vertices 
%(Fig.~\ref{GZsqsqFeyn})
}
\item{The quark- squark- chargino vertices 
%(Fig.~\ref{qsqchiFeyn})
}
\item{The lepton - slepton - chargino vertices 
%(Fig.~\ref{eSneChaFeyn})
}
\end{enumerate}

We begin by explaining the origin of chargino mass eigenstates.
Suppose we write the components of the
$\hat{W}$ superfield and the two Higgs superfields
$\hat{H}_U$ and $\hat{H}_D$ as:
\begin{eqnarray}
\hat{W}^{\pm}=(W_{\mu}^\pm , \lambda^\pm)
\nonumber \\
\hat{H}_U=\left(
\begin{array}{c}
(H_U^0,\tilde{H}_U^0) \\
(H_U^-,\tilde{H}_U^-) 
\end{array}
\right)
\nonumber \\
\hat{H}_D=\left(
\begin{array}{c}
(H_D^+,\tilde{H}_D^+) \\
(H_D^0,\tilde{H}_D^0) 
\end{array}
\right)
\label{supermultiplets}
\end{eqnarray}
Then the chargino matrix is given by:
\begin{eqnarray}
(\lambda^- \ \ \tilde{H}_U^- )
\left(
\begin{array}{cc}
 M    &  \sqrt{2}m_Ws_{\beta}  \\
\sqrt{2}m_Wc_{\beta} & \mu 
\end{array}
\right)
\left(
\begin{array}{c}
\lambda^+ \\
\tilde{H}_D^+
\end{array}
\right)
\label{charginomatrixbasis}
\end{eqnarray}
In a standard notation $M$ is the soft supersymmetry breaking mass
of the winos, $\mu$ is the supersymmetry preserving mass of the Higgsinos
and the mixing between winos and Higgsinos originates from
the supersymmetric version of the $W^\pm H^\pm H^0$ vertex in a two Higgs 
doublet model where the $W^\pm$ and $H^\pm$ are replaced by their
superpartners and the $H^0$ is replaced by its vacuum expectation value.

\begin{figure}
\centerline{\protect\hbox{\psfig{file=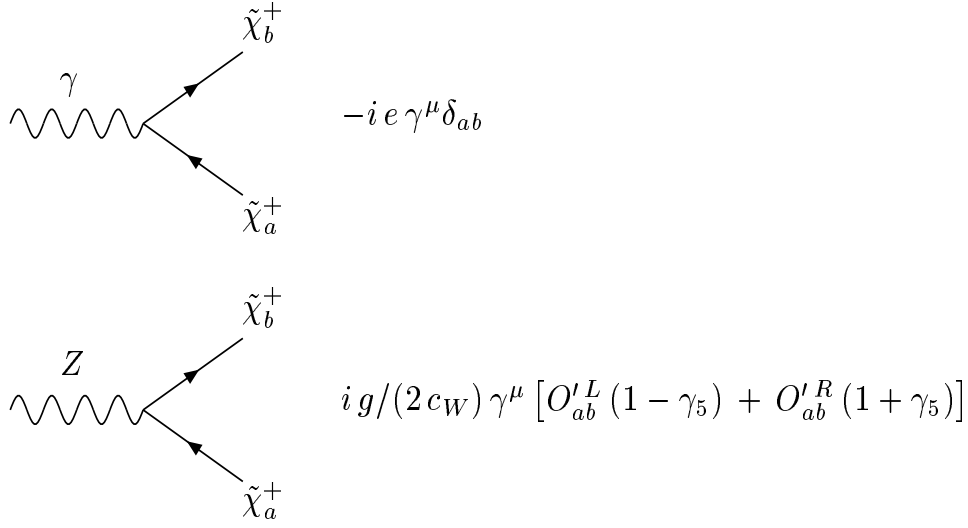,height=7cm,width=0.8\textwidth}}}
\caption{$\gamma\chi^+\chi^+$ and $Z\chi^+\chi^+$ Feynman rules.} 
\label{GZchachaFeyn} 
\end{figure} 
The chargino matrix is diagonalized by two independent
unitary matrices $U$ and $V$:
\begin{eqnarray}
U^{\ast}\left(
\begin{array}{cc}
 M    &  \sqrt{2}m_Ws_{\beta}  \\
\sqrt{2}m_Wc_{\beta} & \mu 
\end{array}
\right)V^{-1}
=\left(
\begin{array}{cc}
 m_{\chi_1} & 0  \\
0 & m_{\chi_2}
\end{array}
\right)
\label{charginomatrix}
\end{eqnarray}
where
\begin{equation}
2m_{\chi_{1,2}}=
M^2+\mu^2+2m_W^2 \\
\mp \sqrt{(M^2-\mu^2)^2+4m_W^4c^2_{2\beta}+
4m_W^2(M^2+\mu^2+2M\mu s_{2\beta})  }
\end{equation}
After diagonalization the four Weyl spinors
$\lambda^+, \lambda^-, \tilde{H}_D^+, \tilde{H}_U^- $
are related to four mass eigenstate Weyl spinors
$\chi_1^+,\chi_2^+,\chi_1^-,\chi_2^-$,
\begin{eqnarray}
\left(
\begin{array}{c}
\chi_1^+ \\
\chi_2^+
\end{array}
\right)
=
V
\left(
\begin{array}{c}
\lambda^+ \\
\tilde{H}_D^+
\end{array}
\right)
\label{Weylcharginos1}
\end{eqnarray}
\begin{eqnarray}
\left(
\begin{array}{c}
\chi_1^- \\
\chi_2^-
\end{array}
\right)
=
U
\left(
\begin{array}{c}
\lambda^- \\
\tilde{H}_U^-
\end{array}
\right)
\label{Weylcharginos2}
\end{eqnarray}
Clearly $\chi_1^+$ and $\chi_1^-$ have a common mass $m_{\chi_1}$.
Similarly $\chi_2^+$ and $\chi_2^-$ have a common mass $m_{\chi_2}$.
In view of this one may define Dirac spinors $\tilde{\chi}_1^+$
and $\tilde{\chi}_2^+$ as:
\begin{eqnarray}
\tilde{\chi}_1^+=\left(
\begin{array}{c}
\chi_1^+ \\
\bar{\chi}_1^-
\end{array}
\right)
\label{Diraccharginos1}
\end{eqnarray}
\begin{eqnarray}
\tilde{\chi}_2^+=\left(
\begin{array}{c}
\chi_2^+ \\
\bar{\chi}_2^-
\end{array}
\right)
\label{Diraccharginos2}
\end{eqnarray}
where $\bar{\chi}_1^-$ and $\bar{\chi}_2^-$ are the CP conjugates
of the Weyl spinors ${\chi}_1^-$ and ${\chi}_2^-$.

When we write down the Feynman rules it will be in terms of these
Dirac mass eigenstates $\tilde{\chi}_1^+$
and $\tilde{\chi}_2^+$ which are four-component spinors
and have antiparticles $\tilde{\chi}_1^-$
and $\tilde{\chi}_2^-$.

In Fig.~\ref{GZchachaFeyn} the photon vertex is standard and the $Z$ 
vertex is in the notation of Haber and Kane where:
\begin{eqnarray}
O^{'L}_{ab} & = & -V_{a1}V^{\ast}_{b1}-\frac{1}{2}V_{a2}V^{\ast}_{b2}
+\delta_{ab}s_W^2 \nonumber \\
O^{'R}_{ab} & = & -U^{\ast}_{a1}U_{b1}-\frac{1}{2}U^{\ast}_{a2}U_{b2}
+\delta_{ab}s_W^2
\end{eqnarray}
where $U$ and $V$ are the matrices which diagonalize the chargino mass
matrix.

\begin{figure}
\centerline{\protect\hbox{\psfig{file=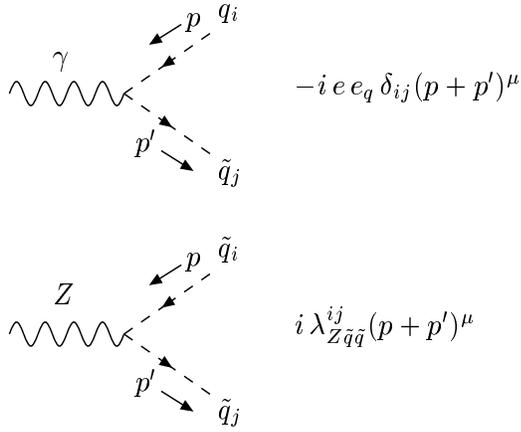,height=7cm,width=0.5\textwidth}}}
\caption{$\gamma\tilde q\tilde q$ and $Z\tilde q\tilde q$ Feynman rules.} 
\label{GZsqsqFeyn} 
\end{figure} 
We now turn to the squark vertices in Fig.~\ref{GZsqsqFeyn}.
The photon coupling is standard, and the $Z$ coupling to
the left and right handed stop and sbottom
$\tilde{t}_L,\tilde{t}_R,\tilde{b}_L,\tilde{b}_R$ are given in Haber and Kane
as:
\begin{eqnarray}
Z\tilde{t}_L\tilde{t}_L & : &  \ \ 
\frac{ig}{2c_W}(-1+2e_us_W^2)(p+p')^{\mu}
\nonumber \\
Z\tilde{t}_R\tilde{t}_R & : &  \ \ 
\frac{ig}{2c_W}(e_us_W^2)(p+p')^{\mu}
\nonumber \\
Z\tilde{b}_L\tilde{b}_L & : &  \ \ 
\frac{ig}{2c_W}(1+2e_ds_W^2)(p+p')^{\mu}
\nonumber \\
Z\tilde{b}_R\tilde{b}_R & : &  \ \ 
\frac{ig}{2c_W}(e_ds_W^2)(p+p')^{\mu}
\end{eqnarray}
where $e_u=2/3,e_d=-1/3$.

The coupling in Fig.~\ref{GZsqsqFeyn} refers to the mass eigenstate 
squarks labelled by indices $i,j$, and involves a diagonalisation of the
stop and sbottom mass matrices:
\begin{eqnarray}
(\tilde{t}^{\ast}_L,\tilde{t}^{\ast}_R)
\left(
\begin{array}{cc}
 M_Q^2+m_t^2+\frac{1}{6}(4m_W^2-m_Z^2)c_{2\beta}   & 
m_t(A_U-\mu/t_{\beta})   \\
m_t(A_U-\mu/t_{\beta}) & 
 M_U^2+m_t^2+\frac{2}{3}(m_Z^2-m_W^2)c_{2\beta} 
\end{array}
\right)
\left(
\begin{array}{c}
\tilde{t}_L \\ 
\tilde{t}_R
\end{array}
\right)
\nonumber\\
\label{stopsbotmatrix}\\
(\tilde{b}^{\ast}_L,\tilde{b}^{\ast}_R)
\left(
\begin{array}{cc}
 M_Q^2+m_b^2-\frac{1}{6}(2m_W^2+m_Z^2)c_{2\beta}   & 
m_b(A_D-\mu t_{\beta})   \\
m_b(A_D-\mu t_{\beta})   &
 M_D^2+m_b^2-\frac{1}{3}(m_Z^2-m_W^2)c_{2\beta} 
\end{array}
\right)
\left(
\begin{array}{c}
\tilde{b}_L \\ 
\tilde{b}_R
\end{array}
\right)\nonumber
\end{eqnarray}

The mass eigenstates 
$\tilde{t}_1,\tilde{t}_2,\tilde{b}_1,\tilde{b}_2$ 
are related to the original states
$\tilde{t}_L,\tilde{t}_R,\tilde{b}_L,\tilde{b}_R$ by
rotations through angles $\alpha_t$ and $\alpha_b$:
\begin{eqnarray}
\left(
\begin{array}{c}
\tilde{t}_L \\ 
\tilde{t}_R
\end{array}
\right)
=
R_{\alpha_t}^{\dagger}
\left(
\begin{array}{c}
\tilde{t}_1 \\ 
\tilde{t}_2
\end{array}
\right)
\label{stoprotation}
\end{eqnarray}
\begin{eqnarray}
\left(
\begin{array}{c}
\tilde{b}_L \\ 
\tilde{b}_R
\end{array}
\right)
=
R_{\alpha_b}^{\dagger}
\left(
\begin{array}{c}
\tilde{b}_1 \\ 
\tilde{b}_2
\end{array}
\right)
\label{sbottomrotation}
\end{eqnarray}
where
\begin{eqnarray}
R_{\alpha}=
\left(
\begin{array}{cc}
 c_{\alpha}    &  s_{\alpha}  \\
- s_{\alpha} & c_{\alpha}  
\end{array}
\right)
\label{Rmatrix}
\end{eqnarray}

{}From the above results we obtain the couplings 
$\lambda^{ij}_{Z\tilde t\tilde t}$ and $\lambda^{ij}_{Z\tilde b\tilde b}$ 
used in Fig.~\ref{GZsqsqFeyn}:
\begin{eqnarray}
\lambda^{ij}_{Z\tilde t\tilde t} & = & R_{\alpha_t}^{i1}
R_{\alpha_t}^{j1}\frac{g}{2c_W}(-1+2e_us_W^2) +   
R_{\alpha_t}^{i2}R_{\alpha_t}^{j2}\frac{g}{2c_W}(e_us_W^2)
\nonumber \\
\lambda^{ij}_{Z\tilde b\tilde b} & = & R_{\alpha_b}^{i1}
R_{\alpha_b}^{j1}\frac{g}{2c_W}(1+2e_ds_W^2) +   
R_{\alpha_b}^{i2}R_{\alpha_b}^{j2}\frac{g}{2c_W}(e_ds_W^2)
\label{ZsqsqCouplings}
\end{eqnarray}

\begin{figure}
\centerline{\protect\hbox{\psfig{file=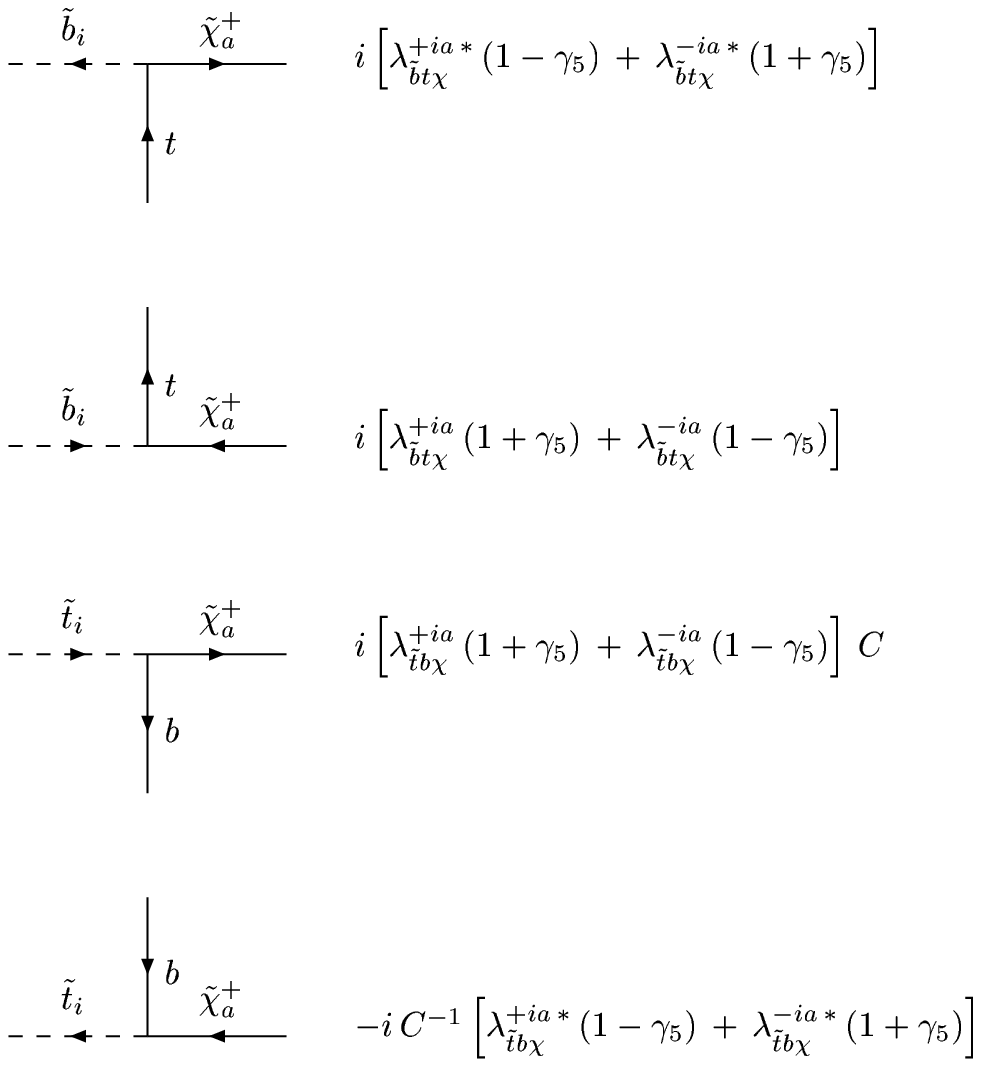,height=14cm,width=0.8\textwidth}}}
\caption{$q\tilde q\chi$ Feynman rules.} 
\label{qsqchiFeyn} 
\end{figure} 
Next we consider the quark-squark-chargino vertices in Fig.~\ref{qsqchiFeyn}.
Again the index $i$ refers to squark mass eigenstates, and the 
index $a$ refers to chargino mass eigenstates.
The Feynman rules for such vertices in the chargino mass
eigenstate basis but involving left and right handed quarks
were given in Gunion and Haber \cite{GunHaber}.
As pointed out by Gunion and Haber, and indicated in Fig.~\ref{qsqchiFeyn},
one must be careful to distinguish between the two cases in which the
arrow on the chargino line enters or leaves the vertex.
In addition one will note the appearance of the charge conjugation
matrix $C$ in the vertices involving the $b$ quark.

To understand these facts consider the standard model
terms which appear in the interaction lagrangian
between the $W^\pm$ and the top and bottom quarks:
$$
gW^+_{\mu}\bar{t}_L\gamma^{\mu}b_L + gW^-_{\mu}\bar{b}_L\gamma^{\mu}t_L 
$$
These two terms may be represented by two diagrams, one involving
the creation of a left handed
top quark whose arrow leaves the vertex, and one 
involving the destruction of a left handed top quark whose arrow
enters the vertex. Note that these two diagrams are {\em not} related
by C or P but simply by Hermitian conjugation.
Now the supersymmetric version of these standard model terms is:
$$
\sqrt{2}g\bar{t}_L\lambda^+\tilde{b}_L + 
\sqrt{2}g\bar{b}_L\lambda^-\tilde{t}_L +
\sqrt{2}g\bar{\lambda^+}t_L\tilde{b}_L^{\ast} + 
\sqrt{2}g\bar{\lambda^-}b_L\tilde{t}_L^{\ast} 
$$
Thus there are  four independent supersymmetric vertices,
corresponding to the $\lambda^\pm$ being created and destroyed at the vertex.
We saw earlier that in terms of the Dirac spinors
a four component mass eigenstate chargino $\tilde{\chi}_a^+$
contains both $\lambda^+$ and the  CP conjugate of $\lambda^-$.
In our convention that $\tilde{\chi}_a^+$ is the particle and
$\tilde{\chi}_a^-$ is the antiparticle;
we must take the CP conjugate of the terms involving $\lambda^-$,
and this leads to the appearance of the $C$ matrix in the Feynman rules
involving $\tilde{\chi}_a^+$, as given in Gunion and Haber.

\begin{figure}
\centerline{\protect\hbox{\psfig{file=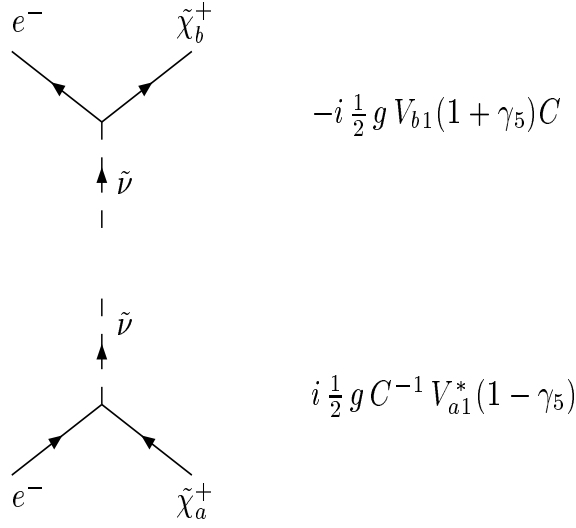,height=7cm,width=0.5\textwidth}}}
\caption{$e^{\pm}\tilde\nu_e\chi^{\pm}$ Feynman rules.} 
\label{eSneChaFeyn} 
\end{figure} 
By replacing the Gunion and Haber left and right handed
quark fields by their mass eigenstates, and correcting 
the sign errors in the $\gamma_5$ matrices in Figs.22(b),
22(d), 23(b) and 23(d)
of ref.\cite{GunHaber} we find that the couplings that we have defined 
in Fig.~\ref{qsqchiFeyn} are given by:
\begin{eqnarray}
\lambda_{\tilde{b}t\chi}^{+ia} & = & -\frac{g}{2}R^{i1}_{\alpha_b}U_{a1}
+\frac{g}{2}\frac{m_b}{\sqrt{2}m_Wc_{\beta}}R^{i2}_{\alpha_b}U_{a2}
\nonumber \\
\lambda_{\tilde{b}t\chi}^{-ia} & = & 
\frac{g}{2}\frac{m_t}{\sqrt{2}m_Ws_{\beta}}R^{i1}_{\alpha_b}V_{a2}^{\ast}
\nonumber \\
\lambda_{\tilde{t}b\chi}^{+ia} & = & -\frac{g}{2}R^{i1}_{\alpha_t}V_{a1}
+\frac{g}{2}\frac{m_t}{\sqrt{2}m_Ws_{\beta}}R^{i2}_{\alpha_t}V_{a2}
\nonumber \\
\lambda_{\tilde{t}b\chi}^{-ia} & = & 
\frac{g}{2}\frac{m_b}{\sqrt{2}m_Wc_{\beta}}R^{i1}_{\alpha_t}U_{a2}^{\ast}
\label{eq:lambda} 
\end{eqnarray}

Finally we note that the Feynman rules in Fig.~\ref{eSneChaFeyn} are 
obtained from Haber and Kane.

\section*{Appendix B: Passarino--Veltman Functions}

\setcounter{equation}{0}
\renewcommand{\theequation}{B.\arabic{equation}}

In this appendix we list all the relevant PV functions, in terms of 
integrals over loop momenta performed in $n=4-2\epsilon$ dimensions.

\subsection*{B1: Tadpole integral}
\begin{equation}
\int \frac{d^nk}{(2\pi)^n} \frac{1}{(k^2-m^2)} =\frac{i}{16\pi^2} A_0(m^2).
\end{equation}

\subsection*{B2: Two-point functions}

\begin{equation}
\int \frac{d^nk}{(2\pi)^n} \frac{1}{(k^2-m_1^2)((k+p)^2-m_2)^2} 
=\frac{i}{16\pi^2} B_0(p^2,m_1^2,m_2^2).
\end{equation}

\begin{equation}
\int \frac{d^nk}{(2\pi)^n} \frac{k^\mu}{(k^2-m_1^2)((k+p)^2-m_2)^2} 
=\frac{i}{16\pi^2} B_1(p^2,m_1^2,m_2^2) \, p^\mu.
\end{equation}

\subsection*{B3: Vertex functions}
\begin{equation}
\int \frac{d^nk}{(2\pi)^n} \frac{1}{(k^2-m_1^2)((k+p_1)^2-m_2)^2
 ((k+p_1+p_2)^2-m_3^2)} 
=\frac{i}{16\pi^2} C_0
\end{equation}

\begin{equation}
\int \frac{d^nk}{(2\pi)^n} \frac{k^\mu}{(k^2-m_1^2)((k+p_1)^2-m_2)^2
 ((k+p_1+p_2)^2-m_3^2)} 
=\frac{i}{16\pi^2} \left( C_{11} \, p_1^\mu+ C_{12} \, p_2^\mu \right),
\end{equation}

\begin{eqnarray} 
\int \frac{d^nk}{(2\pi)^n} \frac{k^\mu k^\nu}{(k^2-m_1^2)((k+p_1)^2-m_2)^2
 ((k+p_1+p_2)^2-m_3^2)} &     \nonumber \\
= \frac{i}{16\pi^2}
\left( C_{21} \, p_1^\mu p_1^\nu + C_{22}\,  p_2^\mu p_2^\nu
 + C_{23} (p_1^\mu p_2^\nu+p_2^\mu p_1^\nu ) + C_{24} g^{\mu\nu} \right),
  & 
\end{eqnarray}
where the functions  $C_0, \, C_{ij}$ have arguments
$(p_1^2,p_2^2,(p_1+p_2)^2,m_1^2,m_2^2,m_3^2)$.   \bigskip

The exact form of the functions $A_0, B_i,C_i,C_{ij}$ are given in \cite{VP}.
The functions $A_0,B_0,B_1,C_{24}$ are ultraviolet divergent and therefore 
contain
the quantity $\Delta$ defined  in eq.~(\ref{eq:Delta}). 
After renormalisation, which corresponds to setting $\Delta=0$, the
functions become finite and are denoted by a tilde, and
depend on the subtraction point $Q$. 
%The 1PI Green's functions given in the following appendix are assumed to 
%have been renormalized. 
For the purposes of our numerical calculations the subtraction
point was taken to be $Q=m_Z$, so that the coupling constants used refer
to running coupings at $m_Z$.

\section*{Appendix C: One-Loop 1PI Green's Functions}

\setcounter{equation}{0}
\renewcommand{\theequation}{C.\arabic{equation}}

In this Appendix we detail our results for the one-loop
Feynman amplitudes obtained using the Feynman rules in Appendix A
and expressed in terms of the PV functions in Appendix B.
Although we shall present results for unrenormalised 1PI Greens
functions, the contributions
to the physical form
factors are obtained from the renormalised 1PI Greens functions.
The renormalised
1PI Greens functions are simply obtained by replacing the singular functions
$A_0,B_0,B_1,C_{24}$ by the corresponding tilded functions as 
explained at the end of the preceeding Appendix. 

\subsection*{C1. Triangular Graphs}

\begin{figure}
\centerline{\protect\hbox{\psfig{file=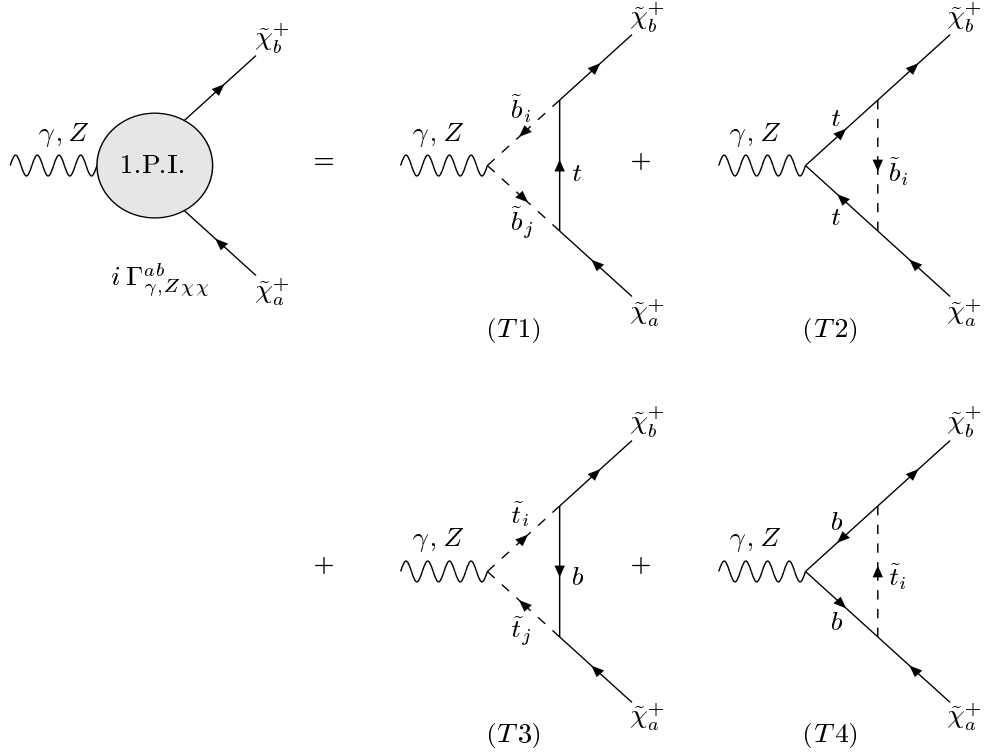,height=10cm,width=0.83\textwidth}}}
\caption{Unrenormalized one--particle irreducible triangular diagrams.} 
\label{Tri1PI} 
\end{figure} 
We begin with the one-loop triangular diagrams contributing to
the Z-chargino-chargino vertex, and displayed in Fig.~\ref{Tri1PI}.

The Z vertex diagram $T1$ in Fig.~\ref{Tri1PI} involving an internal top
quark and two internal sbottom squarks is: 
\begin{eqnarray}
&&\!\!\!\!\!\!\!\!\!\!
[\Gamma_{Z\chi\chi}^{ab}]^{T1}\;\>=\;\>-{{N_c}\over{8\pi^2}}
\sum_{i=1}^2\sum_{j=1}^2
\lambda_{Z\tilde b\tilde b}^{ij}\Bigg\{
\bigg[{\lambda_{\tilde bt\chi}^{+ib}}^*\lambda_{\tilde 
bt\chi}^{+ja}(1-\gamma_5)+{\lambda_{\tilde bt\chi}^{-ib}}^*\lambda_
{\tilde bt\chi}^{-ja}(1+\gamma_5)\bigg]\Big[2C_{24}\gamma^{\mu}
\nonumber\\
&&\quad\qquad\qquad-m_{\chi_a}(2C_{21}+
3C_{11}+C_0)k_1^{\mu}-m_{\chi_a}(2C_{23}+C_{11}+2C_{12}+C_0)k_2^{\mu}\Big]
\label{eq:T1ZXX}\\
&&+m_{\chi_b}\bigg[{\lambda_{\tilde bt\chi}^{+ib}}^*\lambda_{\tilde 
bt\chi}^{+ja}(1+\gamma_5)+{\lambda_{\tilde bt\chi}^{-ib}}^*\lambda_
{\tilde bt\chi}^{-ja}(1-\gamma_5)\bigg]\Big[(2C_{23}+C_{12})k_1^{\mu}+
(2C_{22}+C_{12})k_2^{\mu}\Big]
\nonumber\\
&&-m_t\bigg[{\lambda_{\tilde bt\chi}^{+ib}}^*\lambda_{\tilde 
bt\chi}^{-ja}(1-\gamma_5)+{\lambda_{\tilde bt\chi}^{-ib}}^*\lambda_
{\tilde bt\chi}^{+ja}(1+\gamma_5)\bigg]\Big[(2C_{11}+C_0)k_1^{\mu}+
(2C_{12}+C_0)k_2^{\mu}\Big]\Bigg\}
\nonumber
\end{eqnarray}
where the arguments of the PV functions $C_0$ and $C_{ij}$ are 
$(m_{\chi_a}^2,m_{\chi_b}^2,s;m_{\tilde b_j}^2,m_t^2,m_{\tilde b_i}^2)$.

The Z vertex diagram $T2$ in Fig.~\ref{Tri1PI} involving an internal 
sbottom squark and two internal top quarks is: 
\begin{eqnarray}
&&\!\!\!\!\!\!\!\!\!\!
[\Gamma_{Z\chi\chi}^{ab}]^{T2}\;\>=\;\>{{N_cg}\over{16\pi^2c_W}}
\sum_{i=1}^2\Bigg\{
\nonumber\\
&&\bigg[(g_V^t+g_A^t){\lambda_{\tilde bt\chi}^{+ib}}^*\lambda_
{\tilde bt\chi}^{+ia}(1-\gamma_5)+(g_V^t-g_A^t){\lambda_{\tilde 
bt\chi}^{-ib}}^*\lambda_{\tilde bt\chi}^{-ia}(1+\gamma_5)\bigg]
\bigg[\Big(2C_{24}-B_0^{\chi_b\tilde b_it}-(s
\nonumber\\
&&\quad-m_{\chi_a}^2)C_{12}-m_{\chi_a}^2C_{11}-
m_t^2C_0\Big)\gamma^{\mu}-2m_{\chi_a}(C_{21}+C_{11})k_1^{\mu}
-2m_{\chi_a}(C_{23}+C_{12})k_2^{\mu}\bigg]
\nonumber\\
&&+m_{\chi_b}\bigg[(g_V^t+g_A^t){\lambda_{\tilde bt\chi}^{+ib}}^*
\lambda_{\tilde bt\chi}^{+ia}(1+\gamma_5)+(g_V^t-g_A^t)
{\lambda_{\tilde bt\chi}^{-ib}}^*\lambda_{\tilde bt\chi}^{-ia}
(1-\gamma_5)\bigg]\Big[m_{\chi_a}(C_{12}
\nonumber\\
&&\quad\qquad-C_{11})\gamma^{\mu}+2(C_{23}+C_{12})k_1^{\mu}+2(C_{22}+
C_{12})k_2^{\mu}\Big]
\nonumber\\
&&-m_t\bigg[(g_V^t-g_A^t){\lambda_{\tilde bt\chi}^{+ib}}^*
\lambda_{\tilde bt\chi}^{-ia}(1-\gamma_5)+(g_V^t+g_A^t)
{\lambda_{\tilde bt\chi}^{-ib}}^*\lambda_{\tilde bt\chi}^{+ia}
(1+\gamma_5)\bigg]\Big[m_{\chi_a}C_{11}\gamma^{\mu}
\nonumber\\
&&\quad\qquad-2C_{12}k_2^{\mu}\Big]
\nonumber\\
&&-m_tm_{\chi_b}\bigg[(g_V^t-g_A^t){\lambda_{\tilde bt\chi}^{+ib}}^*
\lambda_{\tilde bt\chi}^{-ia}(1+\gamma_5)+(g_V^t+g_A^t)
{\lambda_{\tilde bt\chi}^{-ib}}^*\lambda_{\tilde bt\chi}^{+ia}
(1-\gamma_5)\bigg]C_{12}\gamma^{\mu}
\nonumber\\
&&+m_t\bigg[(g_V^t+g_A^t){\lambda_{\tilde bt\chi}^{+ib}}^*
\lambda_{\tilde bt\chi}^{-ia}(1-\gamma_5)+(g_V^t-g_A^t)
{\lambda_{\tilde bt\chi}^{-ib}}^*\lambda_{\tilde bt\chi}^{+ia}
(1+\gamma_5)\bigg]\Big[m_{\chi_a}(C_{11}+C_0)\gamma^{\mu}
\nonumber\\
&&\quad\qquad+2(C_{11}+C_0)k_1^{\mu}\Big]
\nonumber\\
&&+m_tm_{\chi_b}\bigg[(g_V^t+g_A^t){\lambda_{\tilde bt\chi}^{+ib}}^*
\lambda_{\tilde bt\chi}^{-ia}(1+\gamma_5)+(g_V^t-g_A^t)
{\lambda_{\tilde bt\chi}^{-ib}}^*\lambda_{\tilde bt\chi}^{+ia}
(1-\gamma_5)\bigg](C_{12}+C_0)\gamma^{\mu}
\nonumber\\
&&+m_t^2\bigg[(g_V^t-g_A^t){\lambda_{\tilde bt\chi}^{+ib}}^*
\lambda_{\tilde bt\chi}^{+ia}(1-\gamma_5)+(g_V^t+g_A^t)
{\lambda_{\tilde bt\chi}^{-ib}}^*\lambda_{\tilde bt\chi}^{-ia}
(1+\gamma_5)\bigg]C_0\gamma^{\mu}\Bigg\}
\label{eq:T2ZXX}
\end{eqnarray}
where the arguments of the PV functions $C_0$ and $C_{ij}$ are 
$(m_{\chi_a}^2,m_{\chi_b}^2,s;m_t^2,m_{\tilde b_i}^2,m_t^2)$,
and $B_0^{\tilde\chi_b\tilde b_i t}\equiv B_0(m_{\chi_b}^2;
m_{\tilde b_i}^2,m_t^2)$.

The Z vertex diagram $T3$ in Fig.~\ref{Tri1PI} involving an internal 
bottom quark and two internal stop squarks is: 
\begin{eqnarray}
&&\!\!\!\!\!\!\!\!\!\!
[\Gamma_{Z\chi\chi}^{ab}]^{T3}\;\>=\;\>-{{N_c}\over{8\pi^2}}
\sum_{i=1}^2\sum_{j=1}^2\lambda_{Z\tilde t\tilde t}^{ij}\Bigg\{
\bigg[\lambda_{\tilde tb\chi}^{-ib}{\lambda_{\tilde 
tb\chi}^{-ja}}^*(1+\gamma_5)+\lambda_{\tilde tb\chi}^{+ib}{\lambda_
{\tilde tb\chi}^{+ja}}^*(1-\gamma_5)\bigg]\Big[2C_{24}\gamma^{\mu}
\nonumber\\
&&\quad\qquad\qquad-m_{\chi_a}(2C_{21}+
3C_{11}+C_0)k_1^{\mu}-m_{\chi_a}(2C_{23}+C_{11}+2C_{12}+C_0)k_2^{\mu}\Big]
\label{eq:T3ZXX}\\
&&+m_{\chi_b}\bigg[\lambda_{\tilde tb\chi}^{-ib}{\lambda_{\tilde 
tb\chi}^{-ja}}^*(1-\gamma_5)+\lambda_{\tilde tb\chi}^{+ib}{\lambda_
{\tilde tb\chi}^{+ja}}^*(1+\gamma_5)\bigg]\Big[(2C_{23}+C_{12})k_1^{\mu}+
(2C_{22}+C_{12})k_2^{\mu}\Big]
\nonumber\\
&&-m_b\bigg[\lambda_{\tilde tb\chi}^{-ib}{\lambda_{\tilde 
tb\chi}^{+ja}}^*(1+\gamma_5)+\lambda_{\tilde tb\chi}^{+ib}{\lambda_
{\tilde tb\chi}^{-ja}}^*(1-\gamma_5)\bigg]\Big[(2C_{11}+C_0)k_1^{\mu}+
(2C_{12}+C_0)k_2^{\mu}\Big]\Bigg\}
\nonumber
\end{eqnarray}
where the arguments of the PV functions $C_0$ and $C_{ij}$ are 
$(m_{\chi_a}^2,m_{\chi_b}^2,s;m_{\tilde t_j}^2,m_b^2,m_{\tilde t_i}^2)$.

The Z vertex diagram $T4$ in Fig.~\ref{Tri1PI} involving internal stop squark
and two internal bottom quarks is: 
\begin{eqnarray}
&&\!\!\!\!\!\!\!\!\!\!
[\Gamma_{Z\chi\chi}^{ab}]^{T4}\;\>=\;\>{{N_cg}\over{16\pi^2c_W}}
\sum_{i=1}^2\Bigg\{
\nonumber\\
&&\bigg[(g_V^b+g_A^b)\lambda_{\tilde tb\chi}^{-ib}{\lambda_
{\tilde tb\chi}^{-ia}}^*(1+\gamma_5)+(g_V^b-g_A^b)\lambda_{\tilde 
tb\chi}^{+ib}{\lambda_{\tilde tb\chi}^{+ia}}^*(1-\gamma_5)\bigg]
\bigg[\Big(2C_{24}-B_0^{\chi_b\tilde t_ib}-(s
\nonumber\\
&&\quad-m_{\chi_a}^2)C_{12}-m_{\chi_a}^2C_{11}-
m_b^2C_0\Big)\gamma^{\mu}-2m_{\chi_a}(C_{21}+C_{11})k_1^{\mu}
-2m_{\chi_a}(C_{23}+C_{12})k_2^{\mu}\bigg]
\nonumber\\
&&+m_{\chi_b}\bigg[(g_V^b+g_A^b)\lambda_{\tilde tb\chi}^{-ib}
{\lambda_{\tilde tb\chi}^{-ia}}^*(1-\gamma_5)+(g_V^b-g_A^b)
\lambda_{\tilde tb\chi}^{+ib}{\lambda_{\tilde tb\chi}^{+ia}}^*
(1+\gamma_5)\bigg]\Big[m_{\chi_a}(C_{12}
\nonumber\\
&&\quad\qquad-C_{11})\gamma^{\mu}+2(C_{23}+C_{12})k_1^{\mu}+2(C_{22}+
C_{12})k_2^{\mu}\Big]
\nonumber\\
&&-m_b\bigg[(g_V^b-g_A^b)\lambda_{\tilde tb\chi}^{-ib}
{\lambda_{\tilde tb\chi}^{+ia}}^*(1+\gamma_5)+(g_V^b+g_A^b)
\lambda_{\tilde tb\chi}^{+ib}{\lambda_{\tilde tb\chi}^{-ia}}^*
(1-\gamma_5)\bigg]\Big[m_{\chi_a}C_{11}\gamma^{\mu}
\nonumber\\
&&\quad\qquad-2C_{12}k_2^{\mu}\Big]
\nonumber\\
&&-m_bm_{\chi_b}\bigg[(g_V^b-g_A^b)\lambda_{\tilde tb\chi}^{-ib}
{\lambda_{\tilde tb\chi}^{+ia}}^*(1-\gamma_5)+(g_V^b+g_A^b)
\lambda_{\tilde tb\chi}^{+ib}{\lambda_{\tilde tb\chi}^{-ia}}^*
(1+\gamma_5)\bigg]C_{12}\gamma^{\mu}
\nonumber\\
&&+m_b\bigg[(g_V^b+g_A^b)\lambda_{\tilde tb\chi}^{-ib}
{\lambda_{\tilde tb\chi}^{+ia}}^*(1+\gamma_5)+(g_V^b-g_A^b)
\lambda_{\tilde tb\chi}^{+ib}{\lambda_{\tilde tb\chi}^{-ia}}^*
(1-\gamma_5)\bigg]\Big[m_{\chi_a}(C_{11}+C_0)\gamma^{\mu}
\nonumber\\
&&\quad\qquad+2(C_{11}+C_0)k_1^{\mu}\Big]
\nonumber\\
&&+m_bm_{\chi_b}\bigg[(g_V^b+g_A^b)\lambda_{\tilde tb\chi}^{-ib}
{\lambda_{\tilde tb\chi}^{+ia}}^*(1-\gamma_5)+(g_V^b-g_A^b)
\lambda_{\tilde tb\chi}^{+ib}{\lambda_{\tilde tb\chi}^{-ia}}^*
(1+\gamma_5)\bigg](C_{12}+C_0)\gamma^{\mu}
\nonumber\\
&&+m_b^2\bigg[(g_V^b-g_A^b)\lambda_{\tilde tb\chi}^{-ib}
{\lambda_{\tilde tb\chi}^{-ia}}^*(1+\gamma_5)+(g_V^b+g_A^b)
\lambda_{\tilde tb\chi}^{+ib}{\lambda_{\tilde tb\chi}^{+ia}}^*
(1-\gamma_5)\bigg]C_0\gamma^{\mu}\Bigg\}
\label{eq:T4ZXX}
\end{eqnarray}
where the arguments of the PV functions $C_0$ and $C_{ij}$ are 
$(m_{\chi_a}^2,m_{\chi_b}^2,s;m_b^2,m_{\tilde t_i}^2,m_b^2)$,
and $B_0^{\tilde\chi_b\tilde t_i b}\equiv B_0(m_{\chi_b}^2;
m_{\tilde t_i}^2,m_b^2)$.

We now turn to the one-loop triangular diagrams contributing to
the photon-chargino-chargino vertex, and displayed in Fig.~\ref{Tri1PI}.

The photon vertex diagram $T1$ in Fig.~\ref{Tri1PI} involving an internal 
top quark and two internal sbottom squarks is: 
\begin{eqnarray}
&&\!\!\!\!\!\!\!\!\!\!
[\Gamma_{\gamma\chi\chi}^{ab}]^{T1}\;\>=\;\>
{{N_cee_b}\over{8\pi^2}}\sum_{i=1}^2\Bigg\{
\bigg[{\lambda_{\tilde bt\chi}^{+ib}}^*\lambda_{\tilde 
bt\chi}^{+ia}(1-\gamma_5)+{\lambda_{\tilde bt\chi}^{-ib}}^*\lambda_
{\tilde bt\chi}^{-ia}(1+\gamma_5)\bigg]\Big[2C_{24}\gamma^{\mu}
\nonumber\\
&&\quad\qquad\qquad-m_{\chi_a}(2C_{21}+
3C_{11}+C_0)k_1^{\mu}-m_{\chi_a}(2C_{23}+C_{11}+2C_{12}+C_0)k_2^{\mu}\Big]
\label{eq:T1PhXX}\\
&&+m_{\chi_b}\bigg[{\lambda_{\tilde bt\chi}^{+ib}}^*\lambda_{\tilde 
bt\chi}^{+ia}(1+\gamma_5)+{\lambda_{\tilde bt\chi}^{-ib}}^*\lambda_
{\tilde bt\chi}^{-ia}(1-\gamma_5)\bigg]\Big[(2C_{23}+C_{12})k_1^{\mu}+
(2C_{22}+C_{12})k_2^{\mu}\Big]
\nonumber\\
&&-m_t\bigg[{\lambda_{\tilde bt\chi}^{+ib}}^*\lambda_{\tilde 
bt\chi}^{-ia}(1-\gamma_5)+{\lambda_{\tilde bt\chi}^{-ib}}^*\lambda_
{\tilde bt\chi}^{+ia}(1+\gamma_5)\bigg]\Big[(2C_{11}+C_0)k_1^{\mu}+
(2C_{12}+C_0)k_2^{\mu}\Big]\Bigg\}
\nonumber
\end{eqnarray}
where the arguments of the PV functions $C_0$ and $C_{ij}$ are 
$(m_{\chi_a}^2,m_{\chi_b}^2,s;m_{\tilde b_i}^2,m_t^2,m_{\tilde b_i}^2)$.

The photon vertex diagram $T2$ in Fig.~\ref{Tri1PI} involving an internal 
sbottom squark and two internal top quarks is: 
\begin{eqnarray}
&&\!\!\!\!\!\!\!\!\!\!
[\Gamma_{\gamma\chi\chi}^{ab}]^{T2}\;\>=\;\>{{N_cee_t}\over{8\pi^2}}
\sum_{i=1}^2\Bigg\{
\bigg[{\lambda_{\tilde bt\chi}^{+ib}}^*\lambda_
{\tilde bt\chi}^{+ia}(1-\gamma_5)+{\lambda_{\tilde 
bt\chi}^{-ib}}^*\lambda_{\tilde bt\chi}^{-ia}(1+\gamma_5)\bigg]
\bigg[\Big(2C_{24}-B_0^{\chi_b\tilde b_it}
\nonumber\\
&&\quad-(s-m_{\chi_a}^2)C_{12}-m_{\chi_a}^2C_{11}
\Big)\gamma^{\mu}-2m_{\chi_a}(C_{21}+C_{11})k_1^{\mu}
-2m_{\chi_a}(C_{23}+C_{12})k_2^{\mu}\bigg]
\nonumber\\
&&+m_{\chi_b}\bigg[{\lambda_{\tilde bt\chi}^{+ib}}^*
\lambda_{\tilde bt\chi}^{+ia}(1+\gamma_5)+
{\lambda_{\tilde bt\chi}^{-ib}}^*\lambda_{\tilde bt\chi}^{-ia}
(1-\gamma_5)\bigg]\Big[m_{\chi_a}(C_{12}-C_{11})\gamma^{\mu}
\nonumber\\
&&\quad\qquad
+2(C_{23}+C_{12})k_1^{\mu}+2(C_{22}+C_{12})k_2^{\mu}\Big]
\nonumber\\
&&+m_t\bigg[{\lambda_{\tilde bt\chi}^{+ib}}^*
\lambda_{\tilde bt\chi}^{-ia}(1-\gamma_5)+
{\lambda_{\tilde bt\chi}^{-ib}}^*\lambda_{\tilde bt\chi}^{+ia}
(1+\gamma_5)\bigg]\Big[m_{\chi_a}C_0\gamma^{\mu}+2(C_{11}+C_0)k_1^{\mu}
\nonumber\\
&&\quad\qquad
+2C_{12}k_2^{\mu}\Big]
\nonumber\\
&&+m_tm_{\chi_b}\bigg[{\lambda_{\tilde bt\chi}^{+ib}}^*
\lambda_{\tilde bt\chi}^{-ia}(1+\gamma_5)+
{\lambda_{\tilde bt\chi}^{-ib}}^*\lambda_{\tilde bt\chi}^{+ia}
(1-\gamma_5)\bigg]C_0\gamma^{\mu}\Bigg\}
\label{eq:T2PhXX}
\end{eqnarray}
where the arguments of the PV functions $C_0$ and $C_{ij}$ are 
$(m_{\chi_a}^2,m_{\chi_b}^2,s;m_t^2,m_{\tilde b_i}^2,m_t^2)$.

The photon vertex diagram $T3$ in Fig.~\ref{Tri1PI} involving an internal 
bottom quark and two internal stop squarks is: 
\begin{eqnarray}
&&\!\!\!\!\!\!\!\!\!\!
[\Gamma_{\gamma\chi\chi}^{ab}]^{T3}\;\>=\;\>
{{N_cee_t}\over{8\pi^2}}\sum_{i=1}^2\Bigg\{
\bigg[\lambda_{\tilde tb\chi}^{-ib}{\lambda_{\tilde 
tb\chi}^{-ia}}^*(1+\gamma_5)+\lambda_{\tilde tb\chi}^{+ib}{\lambda_
{\tilde tb\chi}^{+ia}}^*(1-\gamma_5)\bigg]\Big[2C_{24}\gamma^{\mu}
\nonumber\\
&&\quad\qquad\qquad-m_{\chi_a}(2C_{21}+
3C_{11}+C_0)k_1^{\mu}-m_{\chi_a}(2C_{23}+C_{11}+2C_{12}+C_0)k_2^{\mu}\Big]
\label{eq:T3PhXX}\\
&&+m_{\chi_b}\bigg[\lambda_{\tilde tb\chi}^{-ib}{\lambda_{\tilde 
tb\chi}^{-ia}}^*(1-\gamma_5)+\lambda_{\tilde tb\chi}^{+ib}{\lambda_
{\tilde tb\chi}^{+ia}}^*(1+\gamma_5)\bigg]\Big[(2C_{23}+C_{12})k_1^{\mu}+
(2C_{22}+C_{12})k_2^{\mu}\Big]
\nonumber\\
&&-m_b\bigg[\lambda_{\tilde tb\chi}^{-ib}{\lambda_{\tilde 
tb\chi}^{+ia}}^*(1+\gamma_5)+\lambda_{\tilde tb\chi}^{+ib}{\lambda_
{\tilde tb\chi}^{-ia}}^*(1-\gamma_5)\bigg]\Big[(2C_{11}+C_0)k_1^{\mu}+
(2C_{12}+C_0)k_2^{\mu}\Big]\Bigg\}
\nonumber
\end{eqnarray}
where the arguments of the PV functions $C_0$ and $C_{ij}$ are 
$(m_{\chi_a}^2,m_{\chi_b}^2,s;m_{\tilde t_i}^2,m_b^2,m_{\tilde t_i}^2)$.

The photon vertex diagram $T4$ in Fig.~\ref{Tri1PI} involving internal 
stop squark and two internal bottom quarks is: 
\begin{eqnarray}
&&\!\!\!\!\!\!\!\!\!\!
[\Gamma_{\gamma\chi\chi}^{ab}]^{T4}\;\>=\;\>{{N_cee_b}\over{8\pi^2}}
\sum_{i=1}^2\Bigg\{
\bigg[\lambda_{\tilde tb\chi}^{-ib}{\lambda_
{\tilde tb\chi}^{-ia}}^*(1+\gamma_5)+\lambda_{\tilde 
tb\chi}^{+ib}{\lambda_{\tilde tb\chi}^{+ia}}^*(1-\gamma_5)\bigg]
\bigg[\Big(2C_{24}-B_0^{\chi_b\tilde t_ib}
\nonumber\\
&&\quad-(s-m_{\chi_a}^2)C_{12}-m_{\chi_a}^2C_{11}
\Big)\gamma^{\mu}-2m_{\chi_a}(C_{21}+C_{11})k_1^{\mu}
-2m_{\chi_a}(C_{23}+C_{12})k_2^{\mu}\bigg]
\nonumber\\
&&+m_{\chi_b}\bigg[\lambda_{\tilde tb\chi}^{-ib}
{\lambda_{\tilde tb\chi}^{-ia}}^*(1-\gamma_5)+
\lambda_{\tilde tb\chi}^{+ib}{\lambda_{\tilde tb\chi}^{+ia}}^*
(1+\gamma_5)\bigg]\Big[m_{\chi_a}(C_{12}-C_{11})\gamma^{\mu}
\nonumber\\
&&\quad\qquad
+2(C_{23}+C_{12})k_1^{\mu}+2(C_{22}+C_{12})k_2^{\mu}\Big]
\nonumber\\
&&+m_b\bigg[\lambda_{\tilde tb\chi}^{-ib}
{\lambda_{\tilde tb\chi}^{+ia}}^*(1+\gamma_5)+
\lambda_{\tilde tb\chi}^{+ib}{\lambda_{\tilde tb\chi}^{-ia}}^*
(1-\gamma_5)\bigg]\Big[m_{\chi_a}C_0\gamma^{\mu}+2(C_{11}+C_0)k_1^{\mu}
\nonumber\\
&&\quad\qquad
+2C_{12}k_2^{\mu}\Big]
\nonumber\\
&&+m_bm_{\chi_b}\bigg[\lambda_{\tilde tb\chi}^{-ib}
{\lambda_{\tilde tb\chi}^{+ia}}^*(1-\gamma_5)+
\lambda_{\tilde tb\chi}^{+ib}{\lambda_{\tilde tb\chi}^{-ia}}^*
(1+\gamma_5)\bigg]C_0\gamma^{\mu}
\Bigg\}
\label{eq:T4PhXX}
\end{eqnarray}
where the arguments of the PV functions $C_0$ and $C_{ij}$ are 
$(m_{\chi_a}^2,m_{\chi_b}^2,s;m_b^2,m_{\tilde t_i}^2,m_b^2)$.

\subsection*{C2. Gauge Boson Two--Point Functions}

\begin{figure}
\centerline{\protect\hbox{\psfig{file=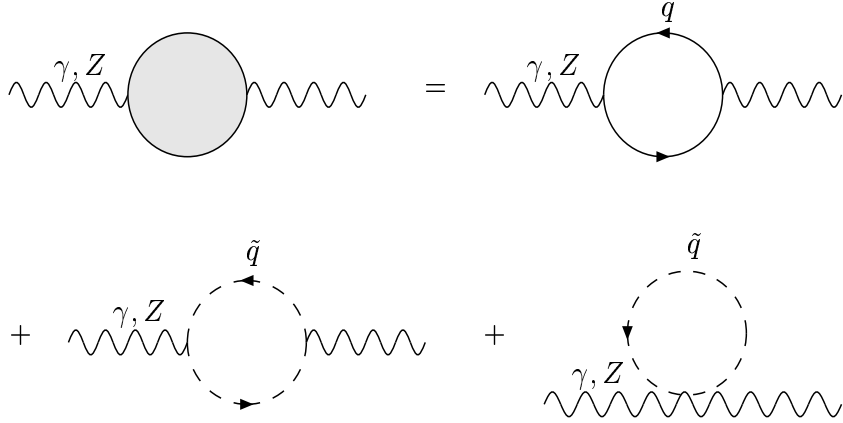,height=7cm,width=0.83\textwidth}}}
\caption{Top and bottom quark and squark contributions to the
unrenormalized self-energy of $\gamma$ or $Z$ } 
\label{ZGse} 
\end{figure} 
The gauge boson two--point functions can be written as $i[A_{GG}(p^2)
g^{\mu\nu}+B_{GG}(p^2)p^{\mu}p^{\nu}]$ where $p$ is the external momentum
and $GG=ZZ$, $\gamma\gamma$, or $Z\gamma$. The functions $A_{GG}$ and 
$B_{GG}$ depend on $p^2$ and represent the one--loop contributions to the 
gauge bosons self energies and mixing. For our purposes only $A_{GG}$ is 
relevant. We detail here the contribution to the gauge bosons self energies 
from top and bottom quarks and squarks, as shown in Fig.~\ref{ZGse}. We 
start with the $Z$--boson self energy. The contribution from top and bottom 
quarks is
\begin{eqnarray}
\Big[A_{ZZ}(p^2)\Big]^{tb}&=&
{{N_cg^2}\over{32\pi^2c_W^2}}(m_t^2B_0^{ptt}+m_b^2B_0^{pbb})
\nonumber\\
&-&{{N_cg^2}\over{16\pi^2c_W^2}}(\quarter-e_ts_W^2+2e_t^2s_W^4)
(4B_{22}^{ptt}-2A_0^t+p^2B_0^{ptt})
\nonumber\\
&-&{{N_cg^2}\over{16\pi^2c_W^2}}(\quarter+e_bs_W^2+2e_b^2s_W^4)
(4B_{22}^{pbb}-2A_0^b+p^2B_0^{pbb})
\label{eq:ZZtb}
\end{eqnarray}
where $A_0^t\equiv A_0(m_t^2)$, $B_{22}^{ptt}\equiv B_{22}
(p^2;m_t^2,m_t^2)$, and similarly for $B_0$. The contribution from
squarks is
\begin{eqnarray}
\Big[A_{ZZ}(p^2)\Big]^{\tilde t\tilde b}&=&
{{N_c}\over{4\pi^2}}\sum_{i=1}^2\sum_{j=1}^2\left[
\lambda_{Z\tilde t\tilde t}^{ij2}B_{22}^{p\tilde t_i\tilde t_j}+
\lambda_{Z\tilde b\tilde b}^{ij2}B_{22}^{p\tilde b_i\tilde b_j}\right]
\nonumber\\
&-&{{N_c}\over{16\pi^2}}\sum_{i=1}^2\left[
\lambda_{ZZ\tilde t\tilde t}^{ij}A_0^{\tilde t_i}+
\lambda_{ZZ\tilde b\tilde b}^{ij}A_0^{\tilde b_i}\right]
\label{eq:ZZstsb}
\end{eqnarray}
where the couplings $\lambda_{Z\tilde t\tilde t}$ and 
$\lambda_{ZZ\tilde t\tilde t}$ are defined in Appendix A.

The photon self energy $A_{\gamma\gamma}$ receive contributions from 
top and bottom quarks:
\begin{equation}
\Big[A_{\gamma\gamma}(p^2)\Big]^{tb}=-{{N_ce^2}\over{8\pi^2}}\bigg[
e_t^2(4B_{22}^{ptt}-2A_0^t+p^2B_0^{ptt})+
e_b^2(4B_{22}^{pbb}-2A_0^b+p^2B_0^{pbb})\bigg]
\label{eq:GGtb}
\end{equation}
and from the top and bottom squarks:
\begin{equation}
\Big[A_{\gamma\gamma}(p^2)\Big]^{\tilde t\tilde b}=
{{N_ce^2}\over{8\pi^2}}\sum_{i=1}^2\bigg[
e_t^2(2B_{22}^{p\tilde t_i\tilde t_i}-A_0^{\tilde t_i})+
e_b^2(2B_{22}^{p\tilde b_i\tilde b_i}-A_0^{\tilde b_i})\bigg]
\label{eq:GGstsb}
\end{equation}
where we take $e$ to be positive.

The top and bottom quarks contributions to the $Z-\gamma$ mixing is
\begin{equation}
\Big[A_{Z\gamma}(p^2)\Big]^{tb}=
-{{N_cge}\over{16\pi^2c_W}}\bigg[e_tg_V^t(4B_{22}^{ptt}-2A_0^t+
p^2B_0^{ptt})+e_bg_V^b(4B_{22}^{pbb}-2A_0^b+p^2B_0^{pbb})\bigg]
\label{eq:ZGtb}
\end{equation}
where $g_V^t=\half(1-4e_ts_W^2)$ and $g_V^b=-\half(1+4e_bs_W^2)$.
In the same way, the contributions from top and bottom squarks can be 
written as
\begin{equation}
\Big[A_{Z\gamma}(p^2)\Big]^{\tilde t\tilde b}=
-{{N_ce}\over{8\pi^2}}\sum_{i=1}^2\bigg[e_t\lambda_{Z\tilde 
t\tilde t}^{ii}(2B_{22}^{p\tilde t_i\tilde t_i}-A_0^{\tilde t_i})+
e_b\lambda_{Z\tilde b\tilde b}^{ii}(2B_{22}^{p\tilde b_i\tilde b_i}-
A_0^{\tilde b_i})\bigg]
\label{eq:ZGstsb}
\end{equation}
and the couplings $Z$--squark--squark are in eq.~\ref{ZsqsqCouplings}.

\subsection*{C3. Chargino Two--Point Functions}

\begin{figure}
\centerline{\protect\hbox{\psfig{file=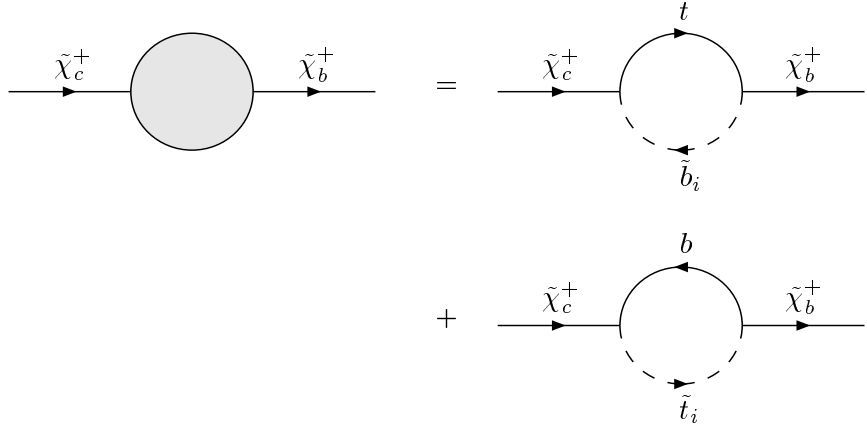,height=7cm,width=0.8\textwidth}}}
\caption{Feynman diagrams contributing to the unrenormalized
chargino two--point functions.} 
\label{fig:cha2point} 
\end{figure} 
In the approximation we are working on, \ie, including only top and 
bottom quarks and squarks in the loops, there are two types of one--loop
graph which contribute to the chargino two--point functions,
and they are displayed in Fig.~\ref{fig:cha2point}.
We use the notation for the sum of the Feynman diagrams
contributing to the chargino two--point functions given in 
eq.~(\ref{eq:ChaChaGeneral}). In this way, the graphs involving top 
quarks and bottom squarks are
\begin{eqnarray}
\left[\Sigma_{\chi\chi}^{ij}(p^2)\right]^{t\tilde b}&=&
i{{N_c}\over{8\pi^2}}\sum_{k=1}^2\Bigg\{\bigg[
{\lambda_{\tilde bt\chi}^{+kj}}^*\lambda_{\tilde bt\chi}^{-ki}(1-\gamma_5)+
{\lambda_{\tilde bt\chi}^{-kj}}^*\lambda_{\tilde bt\chi}^{+ki}(1+\gamma_5)
\bigg]m_tB_0^{pt\tilde b_k}
\nonumber\\ &-&\bigg[
{\lambda_{\tilde bt\chi}^{+kj}}^*\lambda_{\tilde bt\chi}^{+ki}(1-\gamma_5)+
{\lambda_{\tilde bt\chi}^{-kj}}^*\lambda_{\tilde bt\chi}^{-ki}(1+\gamma_5)
\bigg]p_{\mu}\gamma^{\mu}B_1^{pt\tilde b_k}\Bigg\}
\label{eq:ChaChaD1}
\end{eqnarray}
and for bottom quarks and top squarks we have
\begin{eqnarray}
\left[\Sigma_{\chi\chi}^{ij}(p^2)\right]^{b\tilde t}&=&
-i{{N_c}\over{8\pi^2}}\sum_{k=1}^2\Bigg\{\bigg[
\lambda_{\tilde tb\chi}^{-kj}{\lambda_{\tilde tb\chi}^{+ki}}^*(1+\gamma_5)+
\lambda_{\tilde tb\chi}^{+kj}{\lambda_{\tilde tb\chi}^{-ki}}^*(1-\gamma_5)
\bigg]m_bB_0^{pb\tilde t_k}
\nonumber\\ &-&\bigg[
\lambda_{\tilde tb\chi}^{-kj}{\lambda_{\tilde tb\chi}^{-ki}}^*(1+\gamma_5)+
\lambda_{\tilde tb\chi}^{+kj}{\lambda_{\tilde tb\chi}^{+ki}}^*(1-\gamma_5)
\bigg]p_{\mu}\gamma^{\mu}B_1^{pb\tilde t_k}\Bigg\}
\label{eq:ChaChaD2}
\end{eqnarray}
with the couplings given in eq.~(\ref{eq:lambda}).

\end{document}